\RequirePackage{fix-cm}
\documentclass{svjour3}                     
\smartqed  
\usepackage{appendix}
\usepackage{amsmath}
\usepackage{graphicx}
\usepackage{lineno}
\usepackage{array}
\usepackage{longtable}
\usepackage{natbib}

 \usepackage{subfig}
 \usepackage{color}
 \usepackage{multirow}
  \usepackage{tabularx}
  \usepackage{amssymb}
 \usepackage{dashrule}
\usepackage{changes}
 \usepackage[colorinlistoftodos]{todonotes}

\usepackage{booktabs}
\usepackage{slashbox}

\newcounter{commct}
\newcounter{replyct}

 \newcolumntype{P}[1]{>{\centering\arraybackslash}m{#1}}
\newcolumntype{L}[1]{>{\raggedright\let\newline\\\arraybackslash\hspace{0pt}}m{#1}}
\newcolumntype{C}[1]{>{\centering\let\newline\\\arraybackslash\hspace{0pt}}m{#1}}
\newcolumntype{R}[1]{>{\raggedleft\let\newline\\\arraybackslash\hspace{0pt}}m{#1}}
\newcommand{\bs}[1]{\boldsymbol{#1}}

\graphicspath{ {./figs/} }

\newcommand*\patchAmsMathEnvironmentForLineno[1]{%
\expandafter\let\csname old#1\expandafter\endcsname\csname #1\endcsname
\expandafter\let\csname oldend#1\expandafter\endcsname\csname end#1\endcsname
\renewenvironment{#1}%
{\linenomath\csname old#1\endcsname}%
{\csname oldend#1\endcsname\endlinenomath}}%
\newcommand*\patchBothAmsMathEnvironmentsForLineno[1]{%
\patchAmsMathEnvironmentForLineno{#1}%
\patchAmsMathEnvironmentForLineno{#1*}}%
\AtBeginDocument{%
\patchBothAmsMathEnvironmentsForLineno{equation}%
\patchBothAmsMathEnvironmentsForLineno{align}%
\patchBothAmsMathEnvironmentsForLineno{flalign}%
\patchBothAmsMathEnvironmentsForLineno{alignat}%
\patchBothAmsMathEnvironmentsForLineno{gather}%
\patchBothAmsMathEnvironmentsForLineno{multline}%
}

\begin{document}

\title{Prediction of Reynolds Stresses in High-Mach-Number Turbulent Boundary Layers using Physics-Informed Machine Learning
}


\author{Jian-Xun Wang         \and
        Junji Huang           \and 
        Lian Duan             \and 
        Heng Xiao
}


\institute{Jian-Xun Wang \at
              Assistant Professor, Department of Aerospace and Mechanical Engineering; Center of Informatics and Computational Science, University of Notre Dame, Notre Dame, IN, 46556, USA 
             \email{jwang33@nd.ed}           
           \and
           Junji Huang \at
              Graduate Student, Department of Mechanical and Aerospace Engineering, Missouri University of Science and Technology, Rolla, MO, 65409, USA 
              \and
              Lian Duan \at
              Assistant Professor, Department of Mechanical and Aerospace Engineering, Missouri University of Science and Technology, Rolla, MO, 65409, USA 
              \and
              Heng Xiao \at
              Assistant Professor, Kevin T. Crofton Department of Aerospace and Ocean Engineering, Virginia Tech, Blacksburg, VA, 24061, USA
}

\date{Received: DD Month YEAR / Accepted: DD Month YEAR}

\maketitle

\begin{abstract}
Modeled Reynolds stress is a major source of model-form uncertainties in Reynolds-averaged Navier-Stokes (RANS) simulations. Recently, a physics-informed machine-learning (PIML) approach has been proposed for reconstructing the discrepancies in RANS-modeled Reynolds stresses. The merits of the PIML framework has been demonstrated in several canonical incompressible flows. However, its performance on high-Mach-number flows is still not clear. In this work we use the PIML approach to predict the discrepancies in RANS modeled Reynolds stresses in high-Mach-number flat-plate turbulent boundary layers by using an existing DNS database. Specifically, the discrepancy function is first constructed using a DNS training flow and then used to correct RANS-predicted Reynolds stresses under flow conditions different from the DNS. The machine-learning technique is shown to significantly improve RANS-modeled turbulent normal stresses, the turbulent kinetic energy, and the Reynolds-stress anisotropy.  Improvements are consistently observed when different training datasets are used. Moreover, a high-dimensional visualization technique and a distance metrics are used to provide a priori assessment of prediction confidence based only on RANS simulations. This study demonstrates that the PIML approach is a computationally affordable technique for improving the accuracy of RANS-modeled Reynolds stresses for high-Mach-number turbulent flows when there is a lack of experiments and high-fidelity simulations.
\end{abstract}

\section{Introduction}
\label{intro}
\subsection{Significance of Reducing Modeling Discrepancies of Reynolds Stresses}
The information of the full Reynolds-stress tensors of high-speed turbulent flows is of theoretical and practical importance. The physics of Reynolds stresses and their dependence on boundary-layer parameters are critical for the theoretical development of advanced compressibility corrections for Reynolds-Averaged Navier-Stokes (RANS) models~\citep{rumsey2010compressibility}. The information of Reynolds stresses is also useful for generating inflow turbulence for unsteady simulations like direct numerical simulations (DNS) and large eddy simulations (LES) when a synthetic turbulence-generation technique such as the digital-filtering method is used~\citep{Wu17}. The information of Reynolds stresses is commonly obtained from the data of high-fidelity simulations (e.g., DNS or resolved LES) or experimental measurements. However, there are few experimental data at high speeds and most existing DNS and LES are limited to low Reynolds numbers. As a result, the information of the full-field Reynolds-stress tensor in the high-Mach-number and high-Reynolds-number regimes is largely unknown. 

RANS-based models have been widely used for simulating high-Mach-number compressible flows in both academic study and industrial applications~\citep{goldberg2000hypersonic,Smits06,sebastian2016computational}. The information of the Reynolds-stress tensor can be obtained through RANS models with a relatively small computational cost since it is modeled by using local mean flow variables. However, the potentially inaccurate modeling assumptions (e.g., Boussinesq eddy viscosity hypothesis) lead to significant model-form errors in the RANS-modeled Reynolds-stress tensor, which also diminish the predictive capability and accuracy of RANS models, in particular for the flows with curvature, swirl, and strong pressure gradients~\citep{craft1996development}. Recently, significant developments have been made in the data science community, which facilitate the development of data-driven approaches to assist RANS modeling by leveraging the use of existing high-fidelity data sets from both DNS and experiments. Previous efforts in reducing model-form errors in the RANS Reynolds-stress closure have mostly followed parametric approaches, which are based on a generic Bayesian framework proposed by Kennedy and O'Hagan~\citep{k2001bayesian}. Specifically, the model coefficients (e.g., $C_1$ and $C_2$ in the $k$--$\varepsilon$ model) are perturbed and calibrated within the Bayesian framework by assimilating high-fidelity data of the quantities of interest (QoI), e.g., velocity, wall shear stress, and lift coefficient~\citep{oliver2011bayesian,cheung2011bayesian,edeling2014bayesian,edeling2014predictive,ray2016bayesian}. Nonetheless, the parametric approaches are still constrained by assumptions in turbulence models, and the structural model-form errors in the Reynolds stress cannot be captured and corrected. Researchers in turbulence modeling communities have recognized the limitations of parametric approaches and started to reduce the structural model errors locally in the Reynolds stress closure. The past several years have witnessed a few efforts in developing nonparametric data-driven approaches for assisting RANS turbulence modeling~\citep{duraisamy2018turbulence}. For example, Dow and Wang~\citep{dow11quanti} used DNS data to infer the full-field discrepancy in the turbulent eddy viscosity and applied it to a range of channel flows. Duraisamy and co-workers~\citep{parish2016paradigm,singh16using} also improved the RANS model by adding a full-field discrepancy function into the production term of the transport equation of turbulence quantities (e.g., $\omega$) based on experimental data. Although both of them demonstrated that the QoIs, such as velocities and lift coefficients, were improved over the baseline RANS predictions after calibrating discrepancy terms, the Reynolds-stress tensor itself was not shown to be improved. This is because calibrations of discrepancies in the eddy viscosity field or production terms of turbulent transport equations are not able to correct the potentially inaccurate functional relations, e.g., linear eddy viscosity assumption.

\subsection{Reducing Reynolds-Stress Discrepancies: Existing Machine Learning Approaches}
In order to improve the RANS-modeled Reynolds stresses, Xiao and co-workers~\citep{mfu1,mfu2,mfu3} proposed a model-form uncertainty reduction framework that infers the discrepancy of Reynolds-stress tensor in its physical projections by assimilating sparse observation data of velocities. They showed that significant improvement in the predicted velocity field and other QoIs can be achieved by directly calibrating RANS-modeled Reynolds stresses. Improvement is also shown in certain projections of the Reynolds-stresses that are critical to velocity propagation~\citep{mfu1}, while all tensor components of the corrected Reynolds stress were not notably improved. This is due to sparseness of the data and the non-unique mapping between the velocity and the Reynolds-stress tensor. Another limitation of this framework lies in the dependence of the Reynolds stress discrepancy function on the physical coordinates $\mathbf{x}$. As a result, the calibrated discrepancy function was difficult to be extrapolated to flows that use different physical coordinates. Ling et al.~\citep{ling2015evaluation} alleviated the limitation by using a rich set of invariant mean flow features instead of physical coordinates and predicted where the RANS would provide inaccurate results. They also directly predicted the Reynolds stress anisotropy tensor solely based on DNS database by using machine learning algorithms~\citep{ling2016machine,ling2016uncertainty,ling2016reynolds}. Recently, Wang et al.~\citep{Wang2016} proposed a physics-informed, machine-learning (PIML) framework for improving RANS-modeled Reynolds stresses based on a group of training flows with DNS data. Specifically, they employed Random Forest regression to learn the functional form of the discrepancy in RANS-modeled Reynolds-stress tensor with respect to a set of invariant, non-dimensional mean flow features, which are constructed based on local mean flow variables. Since the functional form of RANS-modeled Reynolds stress discrepancy is learned in the mean flow feature space, it can be extrapolated to different flows (at different Reynolds numbers or/and in different geometries) to correct the Reynolds stress predictions where high-fidelity data are not available. They further demonstrated that the ML-corrected Reynolds stresses could lead to an improved velocity field after propagation through RANS equations~\citep{wang2016physics,mfu14,Wu2016}. 

\subsection{Objective and Novelty of the Present Work}
The effectiveness of the PIML approach has been demonstrated in multiple low-speed incompressible flows. This includes scenarios where the training flows have the same geometry as the prediction flow but are different in Reynolds numbers and scenarios where training flows differ from the prediction flow in both Reynolds numbers and geometry. So far, the performance of the PIML approach~\citep{Wang2016} for high-speed, compressible flows has not been fully assessed yet. Given the notable differences between the incompressible flows and high-speed flows, the PIML methods and corresponding conclusions in incompressible flows cannot be extrapolated to high-Mach-number flows without any further investigation. It is not unexpected that the same incompressible mean flow features are not enough for compressible flows, and additional local flow variables such as temperature and density should be considered. Therefore, the objective of the present work is to develop and evaluate the PIML approach by Wang et al.~\citep{Wang2016,mfu14} for predicting Reynolds stresses in the high-speed regime. To fully explore the characteristics of high-speed flows, the mean flow feature space is constructed from the raw mean flow variables by using a systematic approach proposed by Ling et al.~\citep{ling2016machine}. As a first step to extend the PIML framework~\citep{Wang2016} to high-speed compressible flows, high-speed flat-plate turbulent boundary layers are studied in this work. An existing DNS database of turbulent boundary layer flows with different freestream Mach numbers and wall-to-recovery temperature ratios are used to evaluate the learning-prediction performance~\citep{Duan14_Acoustics,Duan14_JFM,Duan16_JFM,zhang2017effect}. The metric of prediction confidence proposed by Wu et al.~\citep{mfu9} and a high-dimensional visualization technique, t-Distributed Stochastic Neighbor Embedding (t-SNE), are employed to analyze the prediction performance with different training flows and to assess the prediction confidence \emph{a priori} based on RANS-predicted mean flow features. It should be noted that the goal of the current study is to infer the Reynolds stress tensor based on RANS, as the knowledge of Reynolds stresses by itself carries important physical values for studying high-speed turbulent flows. The propagation of the predicted Reynolds stresses to other quantities of interest (QoIs) such as the velocity is a separate issue that will be addressed in future studies.

The paper is structured as follows. Section~\ref{sec:meth} outlines the methodology of PIML approach and the construction of mean flow features for compressible flows, including a brief description of the DNS database to be used, the simulation details of the baseline RANS, and the detailed procedures of the PIML approach. Prediction results and discussion of different training sets are shown in Sect.~\ref{sec:results} and~\ref{sec:parameter}, respectively. Summaries are given in Sect.~\ref{sec:con}.

\section{Methodology}
\label{sec:meth}
\subsection{Physics-Informed Machine Learning Approach}
The PIML approach~\citep{Wang2016,mfu14} is that given a set of training flows with data, the functional form of discrepancy $\Delta \mathbf{R}$ in the RANS-modeled Reynolds stress with respect to the RANS-modeled mean flow features $\mathbf{q}$ can be extracted by learning from the training flows. Based on the learned discrepancy function $\Delta \mathbf{R}$, accurate Reynolds stresses of a new flow with a different configuration or a different flow condition can be obtained by performing only RANS simulations. The objective of the PIML approach is to build the mapping between the mean flow to the discrepancy function,
\begin{equation}
f : \mathbf{q} \to \Delta \mathbf{R}.
\end{equation}
There are three essential components in the PIML framework: (1) identification of the mean flow features as the predictors, (2) parameterization of the Reynolds stress discrepancy as the learning target, and (3) construction of the mapping $f$ based on machine learning algorithm and training data. 

Identifying mean flow features is crucial in the PIML approach. The features should be rich enough to reflect the characteristics of each flow. Moreover, to be able to extrapolate among different flows independent of reference frames, the features are required to be dimensionless and invariant. A systematic methodology to construct a complete invariant basis for a given tensorial variables~\citep{ling2016machine,mfu14} is employed to identify the flow features. Specifically, a group of raw mean flow tensors that are critical to the high-speed turbulent flows is chosen first, and a finite integrity basis of invariants for the chosen raw tensors can be constructed by Hilbert basis theorem~\citep{johnson2016handbook}. Since any scalar invariant function of the raw mean flow variables can be expressed as a function of the invariant basis, the features are rich enough to build the Reynolds stress discrepancy by considering high-order and interaction effects. The raw tensorial mean flow variables and normalization scheme chosen in this work are listed in Table~\ref{tab:featureRaw}. The rate-of-strain tensor $\mathbf{S}$, rate-of-rotation tensor $\mathbf{\Omega}$, and gradient of turbulent kinetic energy (TKE)~$\nabla k$ are considered to represent major mean flow physics and to be important to the turbulent constitutive relation. Therefore, these tensorial variables are widely used as building blocks in nonlinear eddy viscosity models~\citep{gatski2000nonlinear,so2004explicit,nieckele2016anisotropic}. The temperature gradient $\nabla T$ is also incorporated as the raw features to account for variations in thermodynamic quantities in compressible flows. Based on these raw mean flow variables, a basis of 47 invariants can be built (see Appendix~\ref{sec:appendx1}). Note that the major difference between features for high-speed flows and those for incompressible flows lies in the inclusion of temperature gradient $\widehat{\nabla T}$ as a raw feature. As a result, the high-order effects of $\widehat{\nabla T}$ and its interactions with other mean flow quantities are considered. The construction of input features is a post-processing step and thus its computational cost is negligible compared to that of the RANS simulation.
\begin{table}[htbp] 
	\def\arraystretch{1.1}		
	\centering
	\caption{
		Raw mean flow feature variables used to construct the integrity invariant basis.
		The normalized feature $\widehat{\alpha}_i$ is obtained by normalizing each element of the corresponding 
		raw input $\alpha_i$ with a normalization factor $\alpha_i^*$ according to $\widehat{\alpha}_i = \alpha_i / (|\alpha_i| + |\alpha_i^*|)$. $\| \cdot \|$ indicate matrix norm.}
	\label{tab:featureRaw}
  \resizebox{\textwidth}{!}{\begin{tabular}{P{2.0cm} | P{3.5cm}  P{3.0cm}  P{5.0cm} }	
		\hline
		normalized raw input $\widehat{\alpha}_i$  & description & raw input $\alpha_i$ &
		normalization factor $\alpha_i^*$  \\ 
		\hline
		$\widehat{\mathbf{S}}$  & strain rate tensor&
		$\mathbf{S}$ & 
		$\dfrac{\varepsilon}{k}$\\  
		\hline
		$\widehat{\bs{\Omega}}$  & rotation rate tensor & $\bs{\Omega}$ &
		$\|\mathbf{\Omega}\|$\\ 
		\hline
		$\widehat{\nabla k}$  & gradient of TKE& $\nabla k$ & $\dfrac{\varepsilon}{\sqrt{k}}$ \\ 
		\hline					 			
		$\widehat{\nabla T}$  & temperature gradient &
		$\nabla T$ & $\|\mathbf{U} \cdot \nabla\mathbf{U}\|/C_p$\\
		\hline					
	\end{tabular}}
\end{table}

The learning target is the discrepancy in the RANS-modeled Reynolds stresses. Since all input features are dimensionless and invariant, the discrepancy terms also need to be non-dimensional and invariant. Therefore, the parameterization of Reynolds stress discrepancy is based on the physical decomposition scheme first proposed by Iaccarino and co-workers~\citep{emory2013modeling} in the context of estimating RANS model-form uncertainties. That is, the Reynolds stress tensor $\mathbf{R}$ is decomposed to its physical projections,
\begin{equation}
\label{eq:tau-decomp}
\boldsymbol{\mathbf{R}} = 2 k \left( \frac{1}{3} \mathbf{I} +  \mathbf{A} \right)
= 2 k \left( \frac{1}{3} \mathbf{I} + \mathbf{V} \Lambda \mathbf{V}^T \right),
\end{equation}
where $k$ is the turbulent kinetic energy; $\mathbf{I}$ is the second order identity matrix; $\mathbf{A}$ is the deviatoric part of $\mathbf{R}$; $\Lambda$ and $\mathbf{V}$ are the orthonormal eigenvalues and eigenvectors of $\mathbf{A}$, respectively.  The discrepancy function $\Delta\mathbf{R}$ is defined based on its six physical projections, which are magnitude $k$, shape $\Lambda$, and orientation $\mathbf{V}$ of the Reynolds stress tensor. Specifically, to impose realizability constraint, the eigenvalues $\Lambda = \textrm{diag}[\lambda_1, \lambda_2, \lambda_3]$ are mapped to the barycentric coordinates [$C_1$, $C_2$, $C_3$], indicating the portion of areas of the three sub-triangles in a Cartesian coordinate $[x_b, y_b]$~\citep{banerjee2007presentation}. As a results, the shape of Reynolds stress anisotropy can be uniquely represented with coordinates $[x_b, y_b]$. To represent the orientation (eigenvectors) of the Reynolds stress, Euler angle system is used, which follows the $z$--$x'$--$z''$ convention in rigid body dynamics~\citep{mfu5}. That is, if a local coordinate system $x$--$y$--$z$ spanned by the three eigenvectors is initially aligned with the global coordinate system ($X$--$Y$--$Z$), the current configuration could be obtained by the following three consecutive intrinsic rotations about the axes of the local coordinate system: (1) a rotation about the $z$ axis by angle $\phi_1$, (2) a rotation about the $x$ axis by $\phi_2$, and (3) another rotation about its $z$ axis by $\phi_3$. Now the RANS Reynolds-stress discrepancy $\Delta R$ can be defined on its physical projections as $\Delta R = [\Delta x_b, \Delta y_b, \Delta \mathrm{Log}k, \Delta\phi]$,
\begin{subequations}
	\label{eq:delta}
	\begin{alignat}{2}
	\Delta \log k & =\log_2 k - \log_2 \tilde{k}^{rans}\\
	\Delta x_b  & = x_b - \tilde{x_b}^{rans}\\
	\Delta y_b & = y_b - \tilde{y_b}^{rans}\\
	\Delta \phi & = \phi - \tilde{\phi}^{rans}, 
	\label{eq:kdelta}    
	\end{alignat}
\end{subequations}
where the superscript $\cdot^{rans}$ indicates RANS computed quantities and discrepancy of the magnitude $\Delta \log k$ is the the logarithm of the ratio of the true $k$ to the RANS-computed $k^{rans}$. The RANS Reynolds stress discrepancy $\Delta R$ is the target to be predicted and the corrected Reynolds stress appears in the RANS equation in its divergence form. Please see Ref.~\citep{Wang2016} for more details. 

With identified mean flow features $\mathbf{q}$ and parameterized Reynolds-stress discrepancy variables $\Delta\mathbf{R}$, a machine learning technique is needed to build the functional relation between $\mathbf{q}$ and $\Delta\mathbf{R}$.  Following the works of Wang et al.~\citep{Wang2016,mfu14}, Random Forest (RF) regression is employed to learn this functional from the training data. The RF model is an ensemble learning technique which aggregates predictions from a number of decision trees. Each decision tree partitions the input feature space to a set of rectangles in a tree-like manner. Namely, the optimization of the tree-based algorithm is to decide which feature variable to split on and what splitting points would be. Assuming that the mean flow feature space is partitioned into $M$ regions $\Omega_m (m = 1 \cdots M)$, the prediction of a single tree at the input point $\mathbf{q}$ is the averaged value $\overline{\Delta R}$ of the region where the input point is located,
\begin{equation}
\Delta R^{tree} = \sum^M_{m=1}{\overline{\Delta R}_m I(\mathbf{q} \in \Omega_m)},
\end{equation}
where $I(\cdot)$ is the indicator function. To determine a (sub)optimal splitting strategy of the feature space in a computationally feasible manner, a greedy algorithm is used to minimize the sum of square errors. We consider to split the $i^{\mathrm{th}}$ component of feature vector $\mathbf{q}$ at splitting point $s$, and the splitting component $\mathbf{q}[i]$ and splitting point $s$ are determined by solving
\begin{equation}
\min_{\mathbf{q}[i], s}\left(\sum^N_{\mathbf{q}^j \in \Omega_1(\mathbf{q}[i], s)}(\Delta R^{j}_d -\overline{\Delta R}_{1;q^j})^2 + \sum^N_{\mathbf{q}^j \in \Omega_2(\mathbf{q}[i], s)}(\Delta R^{j}_d -\overline{\Delta R}_{2;q^j})^2\right),
\end{equation}
where $\Omega_1$ and $\Omega_2$ are two portions of the feature space partitioned by the splitting feature component $\mathbf{q}[i]$ at the splitting point $s$; $\Delta R^{j}_d$ is the $j^{\mathrm{th}}$ training data point obtained from DNS; $N$ is the size of the training dataset. This optimization step is repeated on all of the resulting regions. A major issue of a single decision tree is that it tends to have high variance and less prediction accuracy. This issue can be addressed by aggregating a large number of decision trees, which is the essence of the Random Forest model~\citep{james2013introduction}. In the RF model, the ensemble of trees is built with bootstrap aggregation samples (i.e., sampling with replacement) drawn from the training data. Moreover, only a subset of randomly selected feature components is utilized for constructing each single tree. This strategy reduces the correlation among trees in the ensemble and thus decreases the bias of the ensemble prediction. The computational cost for training the RF model is negligible compared to that of the single RANS simulation. 

\subsection{DNS Database of High-Speed Turbulent Boundary Layers}
The current study focuses on applying the aforementioned physics-informed ML technique to 
high-speed turbulent boundary-layer flows. For this purpose, a DNS database of high-speed turbulent boundary layers is used to train
the functional form of discrepancy in RANS-simulated Reynolds stress. 
Relevant flow conditions of the DNS database are summarized in Table~\ref{tab:FlowCondition_boundary_layer}, which provides the boundary-layer parameters at selected locations where turbulence statistics are gathered. The database includes DNS of spatially-developing turbulent boundary layers over a wide range of freestream Mach numbers ($M_\infty = 2.5 -7.8$). All the DNS cases have a similar Reynolds number of $Re_\tau \approx 400$. 
For each DNS, the full three-dimensional compressible Navier-Stokes equations in conservation form are solved numerically, describing the evolution of the density, momentum, and total energy of the flow.  
An optimized 7th order finite-difference WENO (weighted essentially non-oscillatory) scheme is used to compute the convective flux. Compared with the original finite-difference WENO, introduced by Jiang and Shu~\citep{Jiang96}, the present scheme is optimized by means of limiters~\citep{Taylor07, Wu07} to reduce the numerical dissipation. For viscous flux terms, a 4th-order central difference scheme is used. A 3rd-order low storage Runge-Kutta scheme~\citep{Williamson80} is employed for time integration. This significantly relieves the memory requirement and is well suited for time-accurate simulations like DNS. All the DNS cases fall within the perfect gas regime and the usual constitutive relations for a Newtonian fluid are used:
the viscous stress tensor is linearly related to the rate-of-strain tensor, and the heat flux vector is linearly related to the temperature gradient through Fourier's law. The working fluid is air with viscosity calculated by using Sutherland's law, except for Case M8Tw053 where the working fluid is nitrogen and its viscosity is calculated by using Keyes law~\citep{Keyes51}. Compared to the large differences in boundary-layer properties caused by varying the freestream Mach number and the wall temperature, the differences caused by using a different working fluid in DNS are small, if not negligible. 
A constant molecular Prandtl number of $0.71$ is used for all of the DNS cases. Additional details of the DNS, including the numerical methodology and boundary conditions are described in multiple previous publications~\citep{Duan14_Acoustics,Duan14_JFM,Duan16_JFM,zhang2017effect}. 

\begin{figure}
	\centering
    \subfloat[M6Tw076]{\label{fig:domain_M6} \includegraphics[width=0.4\textwidth]{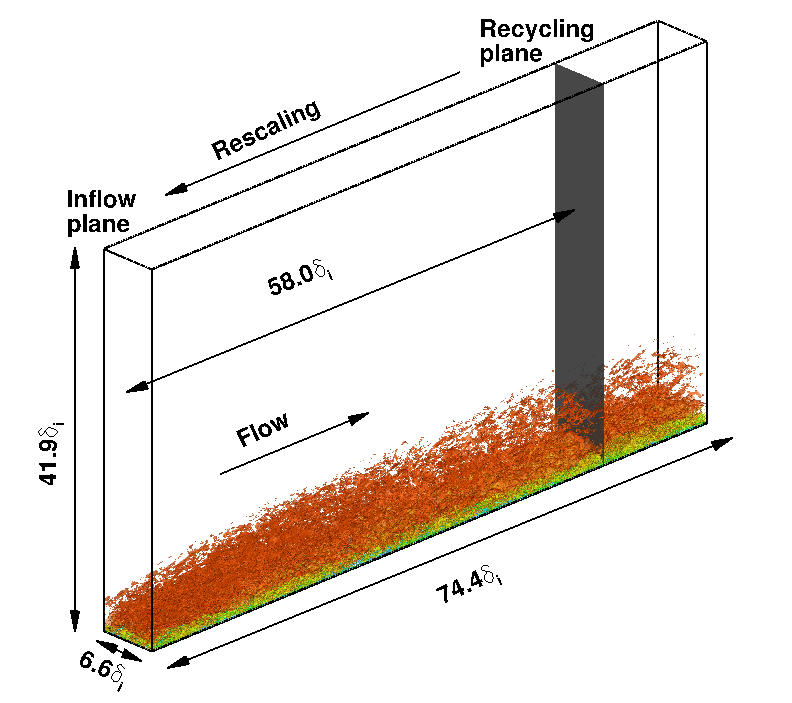}} 
    \subfloat[M8Tw053]{\label{fig:domain_M8} \includegraphics[width=0.53\textwidth]{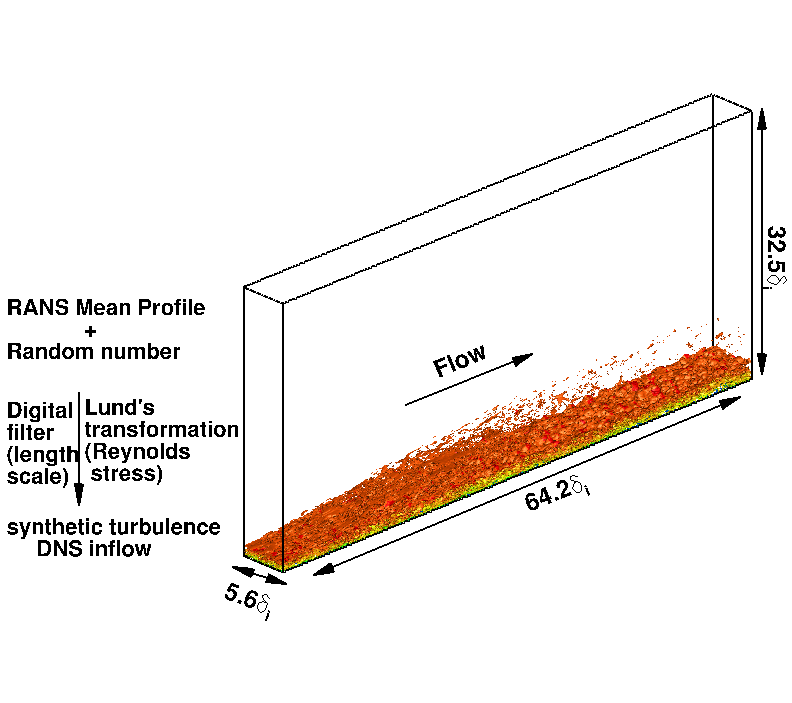}} 
	\caption{Computational domain and simulation setup for DNS of Mach 6 and Mach 8 turbulent boundary layers. The reference length $\delta_i$ is the thickness of the boundary layer (based on $99\%$ of the freestream velocity) at the inlet plane. An instantaneous flow is shown in the domain, visualized by iso-surface of the magnitude of density gradient, $|\Delta \rho|\delta_i/\rho_{\infty} = 0.98$, colored by the streamwise velocity component (with levels from 0 to $U_{\infty}$, blue to red).
    }
	\label{fig:domain_DNS}
\end{figure}

Figs.~\ref{fig:domain_M6} and \ref{fig:domain_M8} show the general computational setup for Cases M6Tw076 and M8Tw053, respectively. 
Inflow boundary condition is prescribed by means of either a recycling-rescaling method adapted from Xu and Martin~\citep{Xu04} or a digital-filtering method introduced by Touber and Sandham~\citep{Touber08}.
The computational setup of the other two cases (M2p5Tw1, M6Tw025) parallels that of Case M6Tw076 in which the rescaling-recycling procedure is used for inflow turbulence generation.
For all cases, a long streamwise domain length is used to minimize any artificial effects of the inflow turbulence generation procedures and to increase the streamwise extent and the Reynolds number range of the DNS.
On the wall, no-slip conditions are applied for the three velocity components and an isothermal condition is used for the temperature. At the outlet and upper freestream boundaries, unsteady non-reflecting boundary conditions based on Thompson~\citep{Thompson87} are imposed to avoid acoustic reflections at the boundaries. Periodic boundary conditions are used in the spanwise direction. Table~\ref{tab:Cases_Grid} summarizes the domain sizes and grid resolutions for various DNS cases. Additional details of the DNS, including the numerical methodology and boundary conditions, are described in multiple previous publications~\citep{Duan14_Acoustics,Duan14_JFM,Duan16_JFM,zhang2017effect}. 

\begin{figure}
	\centering
    \subfloat[]{\label{fig:Mrms_M8} \includegraphics[width=0.3\textwidth]{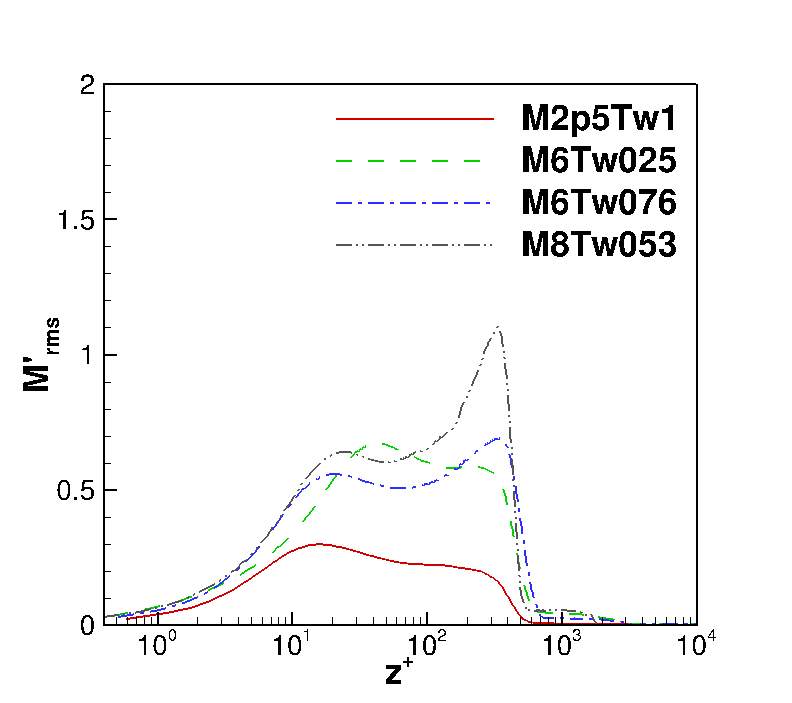}} 
    \subfloat[]{\label{fig:NS_M8} \includegraphics[width=0.66\textwidth]{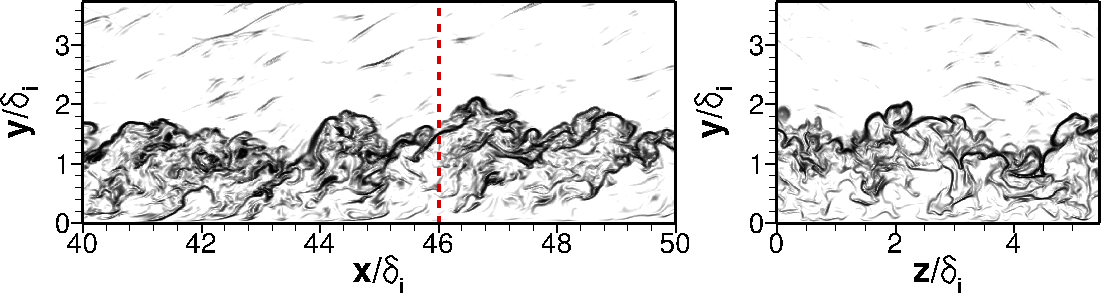}} 
	\caption{(a) Wall-normal distribution of fluctuating Mach number for various Mach number cases, with the wall-normal distance nondimensionalized by wall units, and (b) visualization of a typical instantaneous flow field for DNS Case M8Tw053 in a streamwise wall-normal (x-y) plane, and spanwise wall-normal (z-y) plane. The contours are shown in numerical schlieren, with density gradient contour levels selected to emphasize large scale motions of the boundary layer. The streamwise location of the z-y plane is indicated by the red vertical dashed line.
    }
	\label{fig:compressiblity_M8}
\end{figure}

DNS database covers simulations of turbulent boundary layers in the high-Mach-number, cold-wall regime. The features for such flows can be significantly different from the incompressible counterpart. The different flow features can be caused by changes in mean fluid properties across the layer or due to direct compressibility effects such as the existence of ``eddy shocklets''~\citep{Smits06}. As Mach number increases, the direct compressibility effects become increasingly significant.
For instance, Fig.~\ref{fig:Mrms_M8} shows that the fluctuating Mach number at Mach $7.75$ develops a strong peak with a peak value greater than one toward the edge of the boundary layer. 
As a result, the turbulent fluctuations become locally supersonic relative to the surrounding flow, likely creating local shocklets (see Fig.~\ref{fig:NS_M8}).
Several existing studies have considered compressibility corrections for turbulence models in high-Mach-number boundary-layer applications (see a recent study of Rumsey~\citep{rumsey2010compressibility} and references therein).
However, significant challenges remain for RANS modeling of such highly compressible, shocklet-containing flows, largely due to a lack of Reynolds stress information in such a flow regime. 
In this paper, the PIML approach of Wang et al.~\citep{Wang2016} is used to predict the Reynolds stresses for Case M8Tw053 using one of the other three cases as training flows.
%
%

\begin{table}[h]
	\begin{center}
		\caption{\label{tab:FlowCondition_boundary_layer}
			Boundary layer properties at the center of the domain selected for the analysis for various DNS cases.} 
		\resizebox{\textwidth}{!}{\begin{tabular}{lcccccccccc}
			\hline
			{Case}&{$M_\infty$}&{$T_w/T_r$}&{$ Re_{\theta} $}&{$ Re_{\tau} $}&{$ Re_{\delta_2} $}&{$ \theta (mm)$} & {$ H $} & {$ \delta (mm)$} & {$ z_{\tau} (\mu m)$} &{$ u_{\tau} (m/s)$}\\
			{M2p5Tw1}    &{2.50} &{1.0} &{2187}&{392}&{1280}&{ 0.45 }&{4.0}&{ 5.7 }&{$14.5$} &{ 42.0 }\\ 
			{M6Tw025}&{5.84}&{0.25}&{2028}&{438}&{1086}&{0.19}&{8.4}&{3.5}&{$7.8$} &{34.6}\\ 
			{M6Tw076}&{5.86}&{0.76}&{9818}&{444}&{1910}&{0.95}&{13.5 }&{23.2}&{$52.3$} &{ 45.1 }\\ 
			{M8Tw053}&{7.75}&{0.53}&{8263}&{372}&{1572}&{0.85}&{19.1 }&{25.6}&{$69.3$} &{ 53.5 }\\ 
			\hline
		\end{tabular}}
	\end{center}
\end{table}


\begin{table}[h]
	\begin{center}
		\caption{\label{tab:Cases_Grid} Grid resolutions and domain sizes for direct numerical simulations. 
			$L_x$, $L_y$ and $L_z$ are the domain size in the streamwise, wall-normal, and spanwise directions, respectively. $\Delta x^+$ and $\Delta z^+$ are the uniform grid spacing in the streamwise and spanwise directions, respectively; 
			$\Delta y_{min}^+$ and $\Delta y_{max}^+$ are the minimum and maximum wall-normal grid spacing. 
			The grid resolutions are normalized by the viscous length at the center of the domain selected for the analysis.}
		\resizebox{\textwidth}{!}{\begin{tabular}{lcccccccc}
			\hline
			{Case}&{$N_x \times N_y \times N_z$}&{$L_x/\delta_i$}&{$L_y/\delta_i$}&{$L_z/\delta_i$}&{$\Delta x^+$}&{$\Delta z^+$}&{$\Delta y_{min}^+$}&{$\Delta y_{max}^+$}\\
			{M2p5Tw1}&{$1760\times 400\times 800$}&{57.0}&{40.8}&{15.6}&{9.0}&{5.4}&{0.6}&{9.3}\\
			{M6Tw025}&{$2400\times 560 \times 400$}&{81.5}&{51.1}&{7.9}&{6.5}&{3.8}&{0.5}&{4.8}\\
			{M6Tw076}&{$1920\times500\times320$}&{74.4}&{41.9}&{6.6}&{9.6}&{5.1}&{0.5}&{5.3}\\
			{M8Tw053}&{$3000\times500\times320$}&{64.2}&{32.5}&{5.6}&{4.9}&{3.9}&{0.4}&{3.8}\\
			\hline
		\end{tabular}}
		
	\end{center}
\end{table}

\subsection{Baseline RANS Simulations of High-Speed Turbulent Boundary Layers}
Baseline RANS simulations of 
high-speed turbulent boundary layers are conducted to obtain the mean flow features $\mathbf{q}$ for both the training and 
prediction flows, with the flow conditions and thermodynamic equation of state in RANS matching those of the DNS cases as listed in Table~\ref{tab:FlowCondition_boundary_layer}.
The compressible Favre-averaged Navier-Stokes equations are 
solved using ANSYS Fluent (V15.0)~\citep{guide2013release} with the shear-stress 
transport (SST) $k-\omega$ model of Menter~\citep{menter1994two}.
The SST based $k-\omega$ model differs from the standard $k-\omega$ models in that 
it undergoes a gradual transition to the $k-\epsilon$ model in the outer part of the boundary layer. 
No low-Reynolds-number correction is used as the $k-\omega$ based model can be directly integrated from the wall. Details of the Favre-averaged Reynolds stress equation, boundary conditions, and compressible-flow closure approximations are given in Ref.~\citep{guide2013release}.


\begin{figure}
	\centering
	\includegraphics[width=0.7\linewidth]{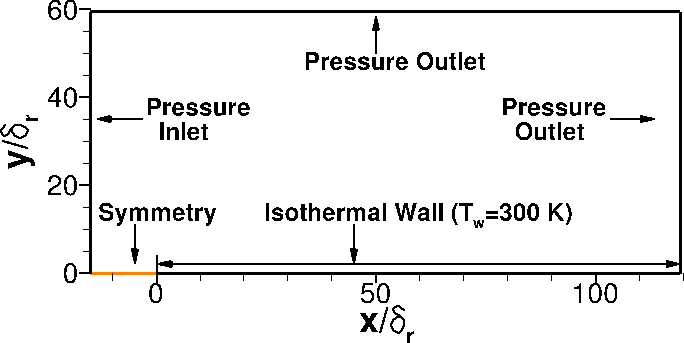}
	\caption{Computation domain and boundary conditions for RANS of a Mach 6 turbulent boundary layer (Case M6Tw076). $\delta_r$ is approximately the boundary-layer thickness at the center of the domain.}
	\label{fig:domain_RANS_M6Tw076}
\end{figure}



Fig.~\ref{fig:domain_RANS_M6Tw076} shows a schematic of the RANS computational domain for Case M6Tw076 along with the boundary conditions used in the Fluent solver. 
A mesh of $561\times150$ grid points is used, respectively, in the streamwise and wall-normal directions. The streamwise and wall-normal domain sizes are approximately $L_x/\delta_r\times L_y/\delta_r = 120\times60$, respectively, where $\delta_r$ is the approximate boundary-layer thickness at the center of the domain.
Uniform grids are used in the streamwise direction with a resolution of $\Delta x/\delta_r \approx 0.3$.  Geometric grids with a stretching ratio of less than $1.05$ are used in the wall-normal direction. The wall-normal grid resolution is $\Delta y^+ \approx 0.8$ at the wall and $\Delta y^+ \approx 16$ near the boundary-layer edge . Systematic grid refinement in each direction has been conducted to verify the grid convergence of the RANS results. The computational setup for RANS of other cases parallels that of 
the Case M6Tw076. 

\section{Learning-Prediction Results}
\label{sec:results}
We focus on demonstrating the physics-informed ML framework on cases where training and prediction flows have the same geometry of a flat plate (with different spatial domain sizes) but are different in flow conditions, including the freestream Mach number $M_{\infty}$ and the wall-to-recovery temperature ratio $T_w/T_r$. Here we demonstrate the effectiveness of the method by presenting a scenario where the Reynolds stresses for Case M8Tw053 are predicted using Case M2p5Tw1 as the training flow. Note that the prediction and training flows have the largest discrepancy in freestream Mach number among all cases considered. Therefore, this scenario is expected to be the most challenging and can best demonstrate the effectiveness of the PIML approach. The Reynolds-stress discrepancy of the lower Mach number training flow (Case M2p5Tw1) is first obtained by comparing the predictions of DNS and RANS for this case. The data is then used to train the Random Forest as the discrepancy function, which is in turn used to correct the RANS-predicted Reynolds stresses for the higher Mach number prediction flow (Case M8Tw053). The ML-assisted RANS results are tested and validated against those of DNS at the condition of the prediction flow. The Random Forest (RF) regressor is constructed of decision trees, which is built to its maximum depth by the successive splitting of nodes until each leaf is left with one training data point. Two free parameters are required for building the RF model, i.e., number of trees and number of randomly selected feature components, on which the split is determined. Generally, a larger ensemble size leads to a better performance. Based on our testing, an ensemble of $200$ trees is large enough to have a robust prediction. The size of the randomly selected subset of features is commonly chosen as the square root of the total number of input features~\citep{friedman2001elements}. In all test cases, the prediction results were shown to be insensitive to this number. The RF model is implemented by using scikit-learn package~\citep{scikit-learn}, which is an open source python library for machine learning tools. In the training process of the RF model, it is not necessary to perform k-fold cross-validation (CV). This is because the bootstrap sampling used in building the RF model enables a simpler way to estimate the test error from training data. Specifically, each tree is trained by using around 2/3 of the data, and the remaining 1/3 can be used to estimate the test error during training. The estimated error is referred to as the out-of-bag (OOB) error, which has been demonstrated to be equivalent to the leave-one-out CV error~\citep{james2013introduction}.
\begin{figure}[htbp]
	\centering
	\includegraphics[width=0.7\linewidth]{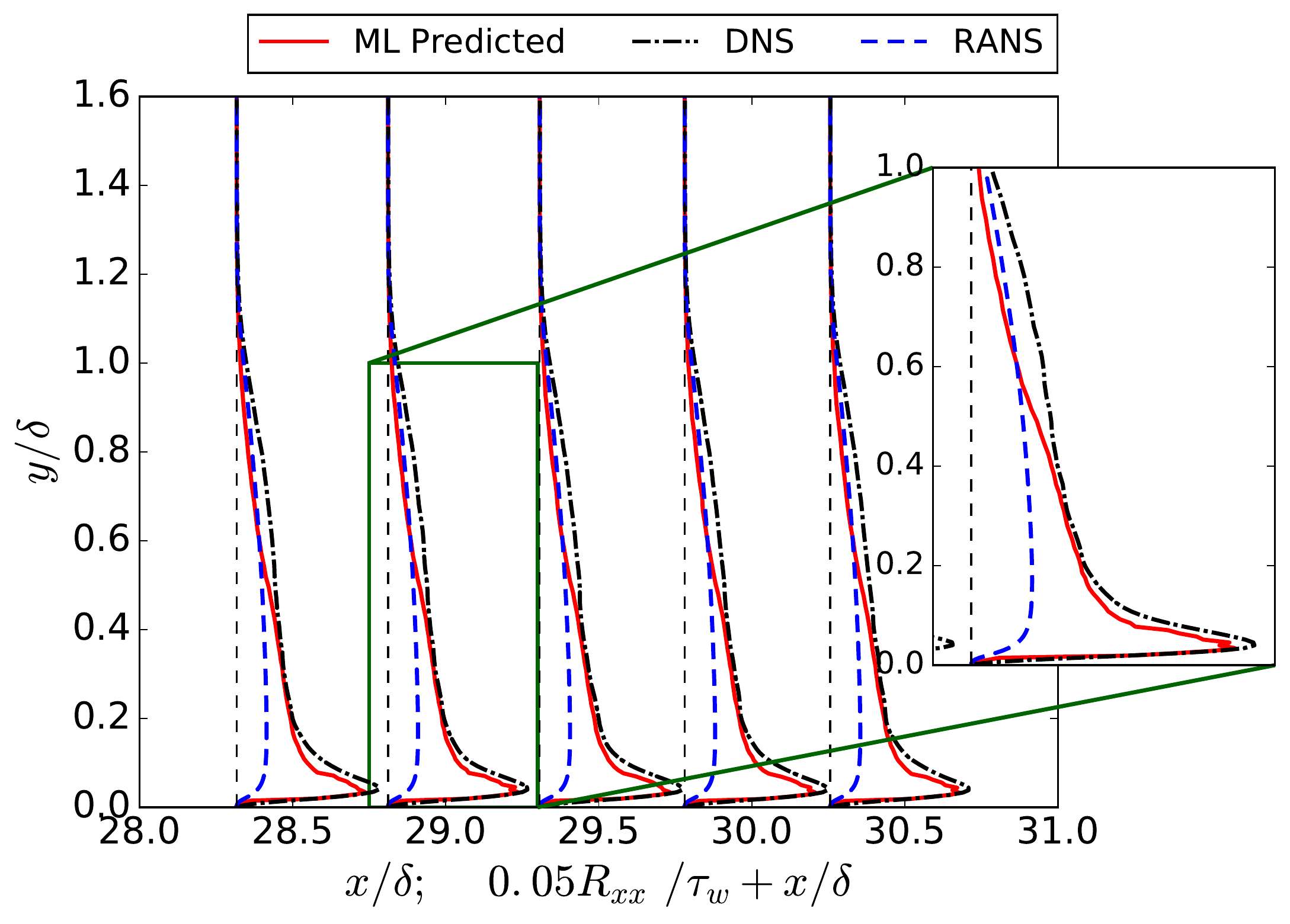}
	\caption{ML Predicted turbulent normal stress $R_{xx} = \overline{\rho u' u'}$ of 
		the test flow (M8Tw053) learned from the training flow (M2p5Tw1) at a low Mach number.
		The profiles are shown at five streamwise locations $x/\delta = 28.32, 28.81, 29.31, 
		29.78, 30.26$. Corresponding baseline RANS predictions and DNS results are
		also plotted for comparison. A zoomed-in view of the profiles at $x/\delta = 28.81$ is 
		presented as an inset to show the detailed comparison.}
	\label{fig:R_xx_M25}
\end{figure}

Fig.~\ref{fig:R_xx_M25} shows the profiles of ML-predicted turbulent normal stress $R_{xx}$ at $x/\delta = 28$ -- $31$,  where the turbulent boundary layer is fully developed. The baseline RANS (i.e. RANS without correction) and the corresponding DNS are also plotted for comparison. It can be seen that the baseline RANS underpredicts the turbulent normal stresses, and there are notable discrepancies in regions close to the wall due to errors in RANS modeling. In contrast, the results corrected by the PIML approach are significantly improved and better agreement with the DNS data is achieved. In particular, the ML predictions capture the sharp peaks near the wall at $y/\delta = 0.05$ for boundary-layer profiles at all the five streamwise locations. For $y/\delta < 0.05$ where turbulence productions are the most significant, the ML-predicted $R_{xx}$ almost overlaps with the DNS data. For $y/\delta > 0.5$, the improvement of ML prediction becomes less notable due to smaller discrepancies in the Reynold stress between the baseline RANS and the DNS. Overall, the ML-predicted turbulent normal stress captures the corresponding DNS data well and significantly improves over the baseline, demonstrating an excellent performance of the physics-informed ML approach. Similarly good agreement is achieved for the other two normal components $R_{yy}$ and $R_{zz}$.

\begin{figure}[htbp]
	\centering
	\includegraphics[width=0.7\linewidth]{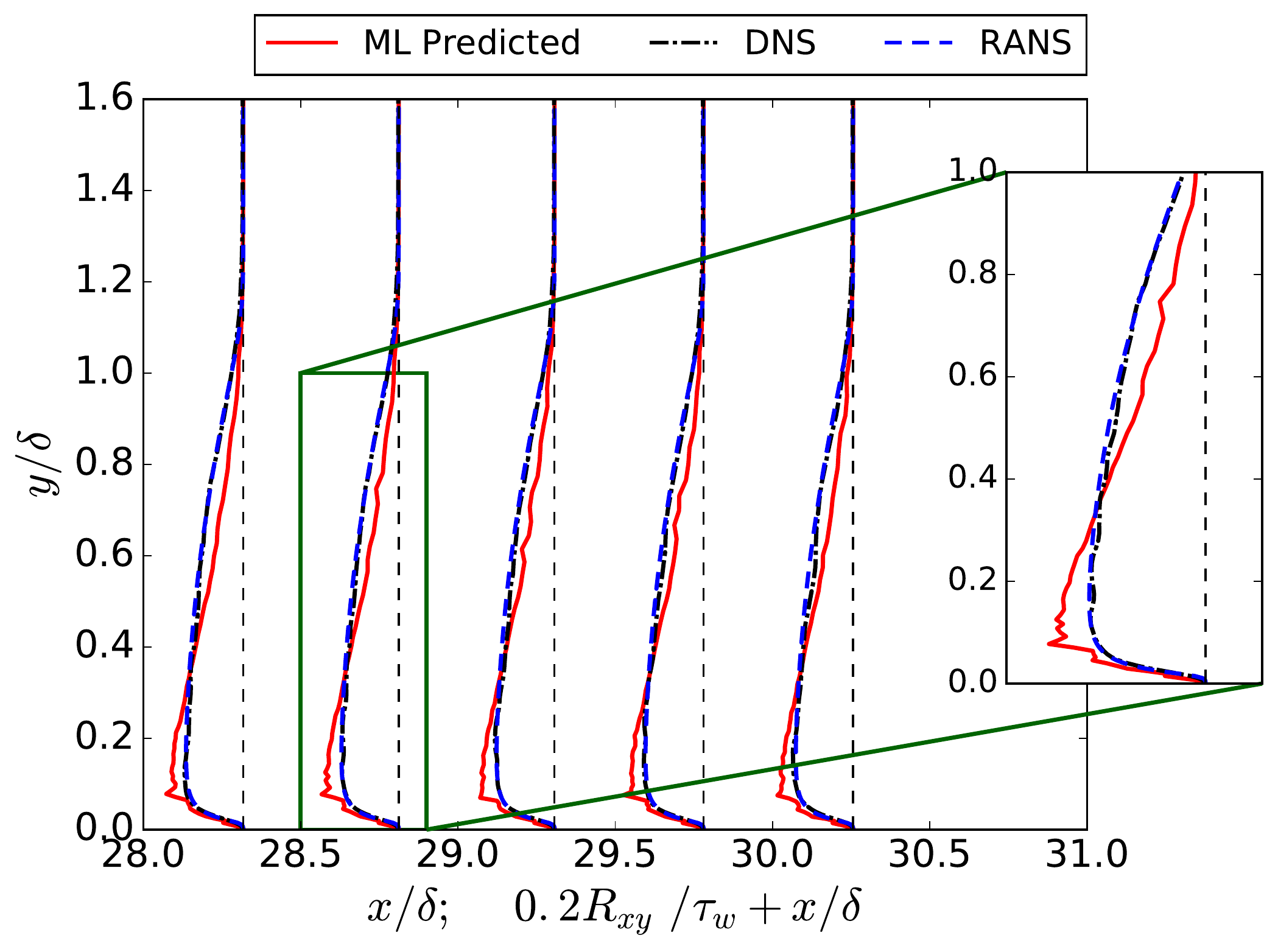}
	\caption{ML Predicted turbulent shear stress $R_{xy} = \overline{\rho u' v'}$ of 
		the test flow (M8Tw053) learned from the training flow (M2p5Tw1) at a low Mach number.
		The profiles are shown at five streamwise locations $x/\delta = 28.32, 28.81, 29.31, 
		29.78, 30.26$. Corresponding baseline RANS predictions and DNS results are
		also plotted for comparison. A zoomed-in view of the profiles at $x/\delta = 28.81$ is 
		presented as an inset to show the detailed comparison.  }
	\label{fig:R_xy_M25}
\end{figure}
 Fig.~\ref{fig:R_xy_M25} shows a comparison of profiles of turbulent shear stress $R_{xy}$ between ML-corrected RANS and the DNS. There is no obvious improvement for the ML-predicted shear stress $R_{xy}$. The limited improvement in $R_{xy}$ by PIML is not unexpected, given that Menter's SST $k$--$\omega$ model used in the baseline RANS has already been well-tuned to provide good predictions of the Reynolds shear stress for canonical flows like attached flat-plate turbulent boundary layers. The minor unsmoothness in some regions of the ML-predicted profile is caused by pointwise estimation of Reynolds stresses in the employed Random Forest algorithm. Such a finding suggests that the PIML methodology needs only to be applied if necessary and a priori understanding of the pro and cons of the employed RANS model would be beneficial.

\begin{figure}[htbp]
	\centering
	 \includegraphics[width=0.45\textwidth]{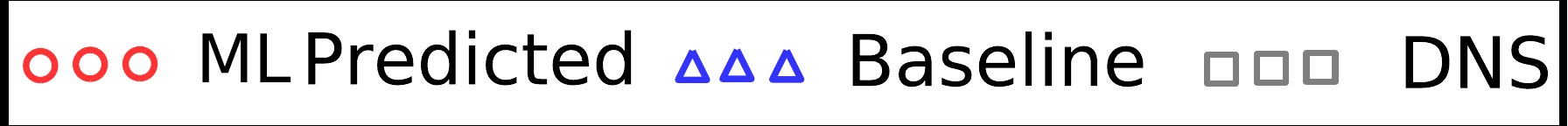}\\
	\subfloat[$x/\delta = 28.81$]{\includegraphics[width=0.45\textwidth]{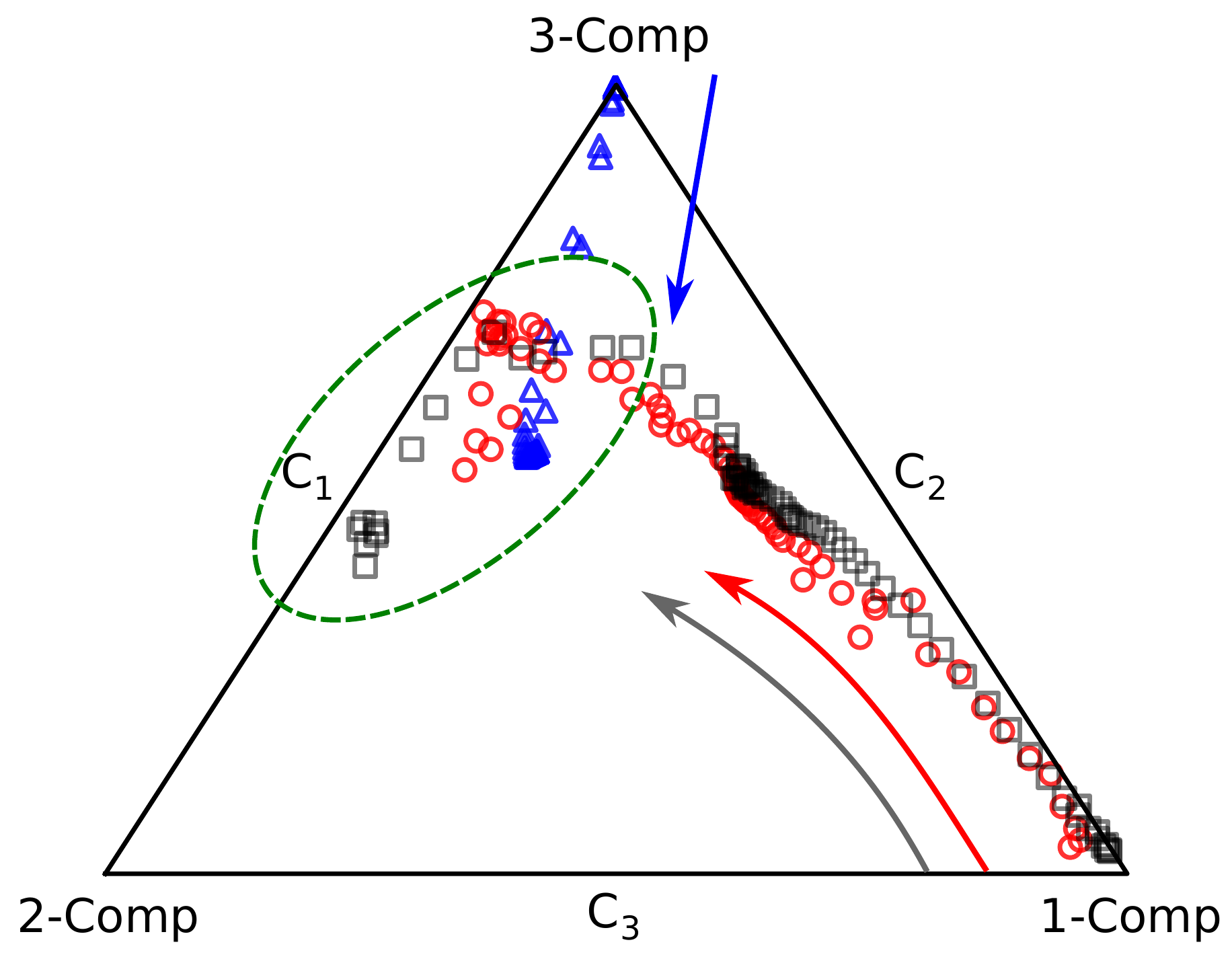}} 
	\subfloat[$x/\delta = 29.78$]{\includegraphics[width=0.45\textwidth]{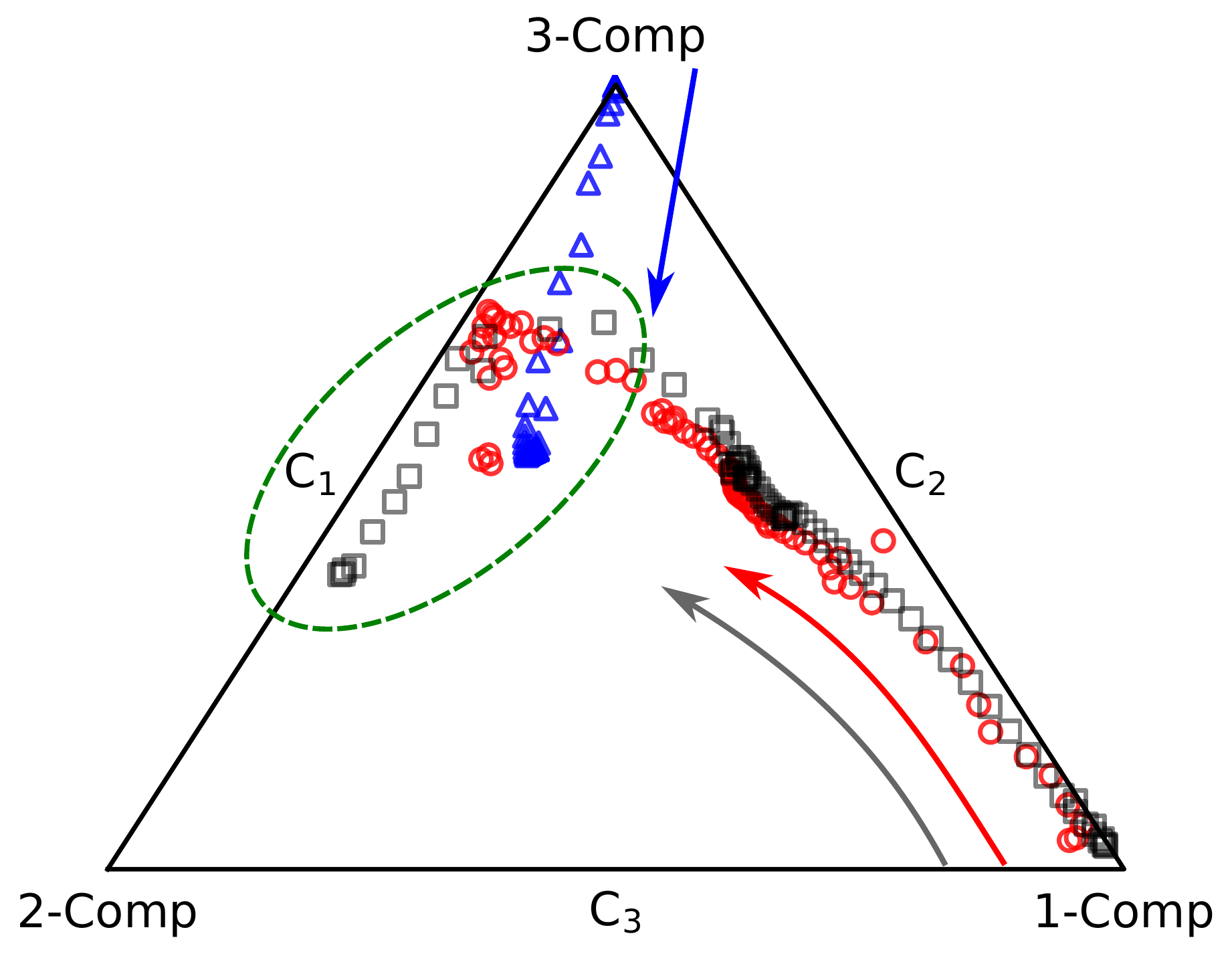}} 		  
	\caption{Barycentric map of the ML corrected Reynolds stress of the test flow (M8Tw053) 
	learned from the training flow (M2p5Tw1) at a low Mach number. The corrected results 
	(red circle) on two streamwise locations (i.e., $x/\delta = 28.81$ and $29.78$) are plotted in 
	panels (a) and (b), respectively. Corresponding RANS baseline results (triangle) and DNS results 
	(square) are also plotted for comparison.}
	\label{fig:bayRe_S2}
\end{figure}

\begin{figure}[htbp]
	\centering
	\includegraphics[width=0.7\linewidth]{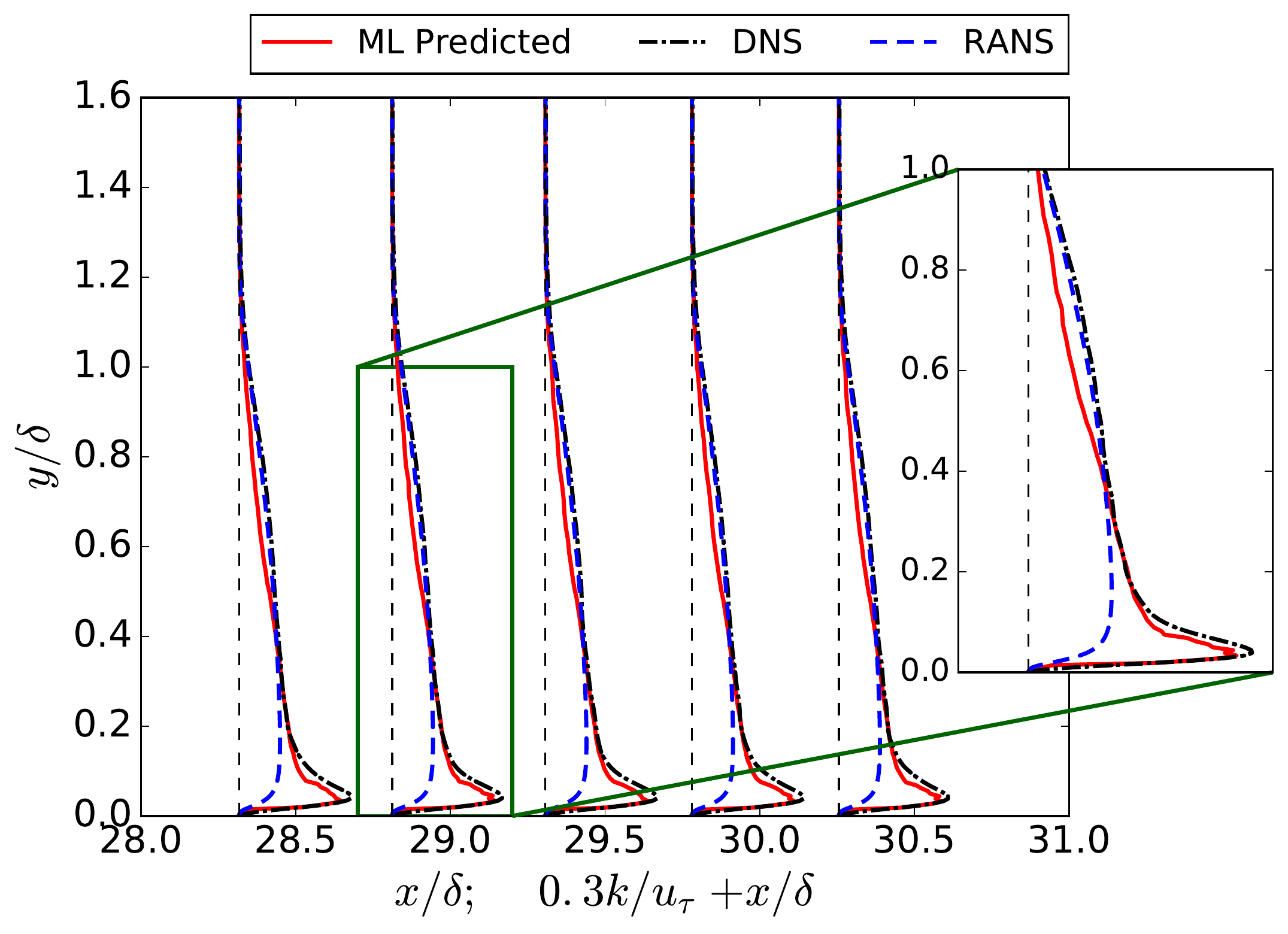}
	\caption{Turbulence kinetic energy ($k$) of the ML Predicted Reynolds-stress tensor normalized
		by the friction velocity $u_{\tau}$. The flow used for training is the one at the Mach number $M = 2.5$ 
		(M2p5Tw1). The profiles are shown at five streamwise locations $x/\delta = 28.32, 28.81, 29.31, 
		29.78, 30.26$. Corresponding baseline RANS and DNS profiles are also plotted for comparison. 
		A zoomed-in view of the profiles at $x/\delta = 28.81$ is 
		presented as an inset to show the detailed comparison. }
	\label{fig:K_M25}
\end{figure}

In addition to the tensor components of Reynolds stresses, it is also of interest to investigate the performance of the ML-predicted Reynolds stress in its physical projections, i.e., Reynolds stress anisotropy and turbulence kinetic energy (TKE). The shape of the anisotropy tensor can be plotted in the barycentric triangle by using barycentric coordinates \citep{banerjee2007presentation}, which provides a non-distorted visual representation of anisotropy due to its linear treatment. Fig.~\ref{fig:bayRe_S2} shows the ML-predicted anisotropy in barycentric triangle compared with the baseline and DNS results. The comparisons are performed on two representative lines at $x/\delta = 28.81$ and $29.78$ and are presented in Figs.~\ref{fig:bayRe_S2}a and~\ref{fig:bayRe_S2}b, respectively. The arrows denote the order of sample points plotted in the triangle, which is from the bottom wall to the outer layer. It shows that the baseline RANS-modeled anisotropy is markedly different from the DNS result since the linear eddy viscosity model (used as the baseline) cannot capture the Reynolds stress anisotropy at all. For both lines, the RANS-modeled Reynolds stresses fall on the plane strain line in the Barycentric triangle, whereas the DNS turbulence states have a significantly different trend. The DNS Reynolds stresses near the wall are close to the one-component limiting state (1-comp, right bottom vertex of the triangle) since the turbulence in the other two directions are dominated by streamwise velocities. In stark contrast, the baseline RANS predicts nearly three-component isotropic state (3-comp, top vertex of the triangle) at the wall. The trend of spatial variations of RANS-predicted anisotropy is opposite to the DNS data. By correcting the baseline RANS results with the trained discrepancy function, the predicted anisotropy of Reynolds stress is significantly improved. For both lines, the predicted anisotropy (circles) agrees well with the DNS result (squares). Moreover, the spatial variations of anisotropy show the same trend as the DNS. It should be noted that the ML-predicted turbulence states enclosed by a dashed ellipse do not well agree with the DNS data though they are significantly improved over the baselines. This is because these turbulence states are located in the regions near or outside the edge of the boundary layer ($y/\delta > 1.0$), where the turbulence almost vanishes (see Fig.~\ref{fig:K_M25}), and thus the anisotropies are difficult to capture.

Fig. 6 shows the comparison in TKE. The TKE predicted by the baseline RANS model has a large discrepancy compared to the DNS, especially in the near-wall region. Although it is well known that the $k$-$\omega$ RANS model alters the physical TKE equation and introduces an extra dissipation near the wall~\citep{durbin2011statistical} that results a much reduced near-wall peak, such a large deviation can be adequately corrected by PIML, indicating the robustness of the employed machine-learning algorithm.

\section{Prediction Performance of Different Training Databases}
\label{sec:parameter}

\begin{figure}[htbp]
	\centering
	\includegraphics[width=0.7\linewidth]{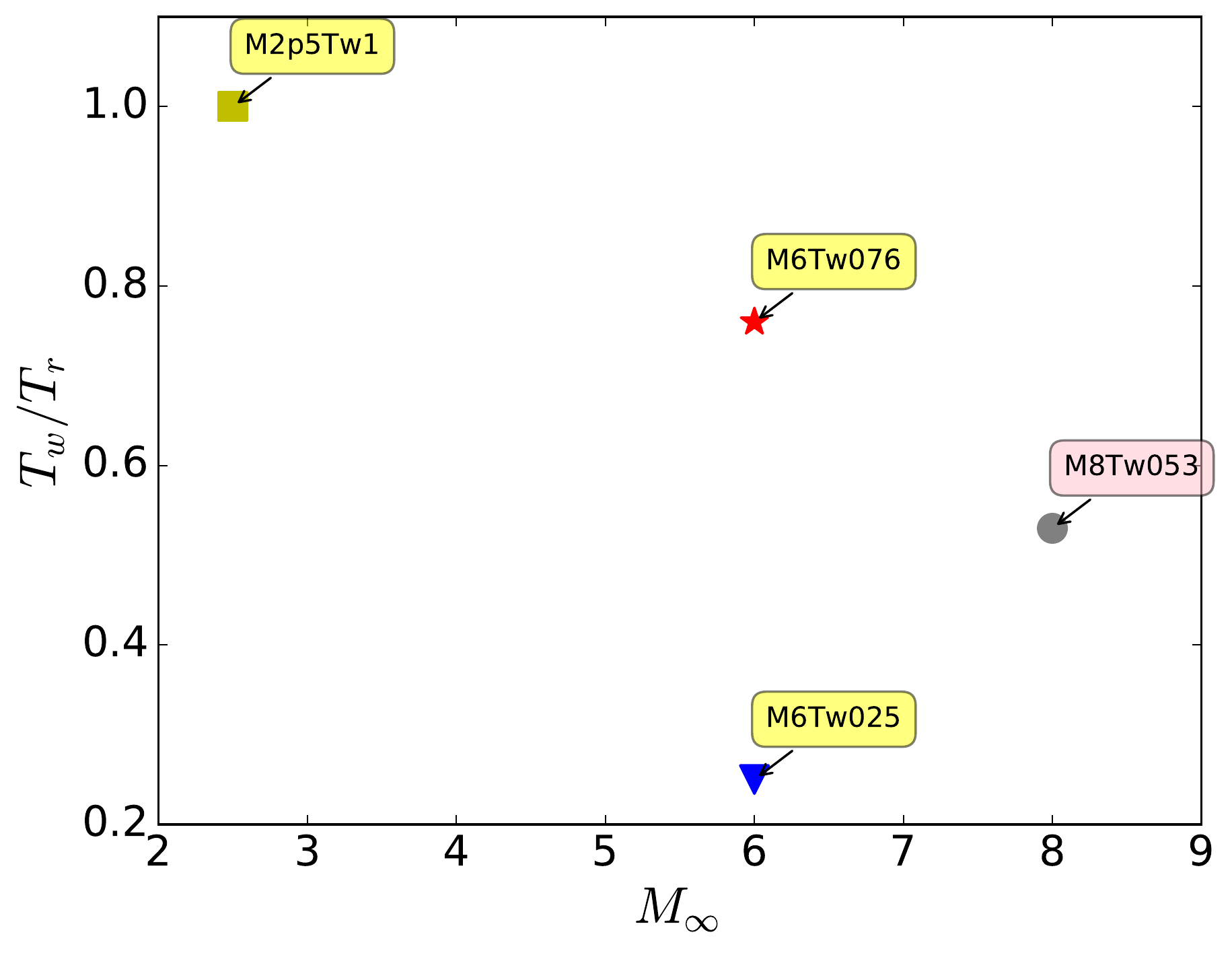}
	\caption{Scatter plots of the four cases in the two-dimensional space of flow conditions, freestream Mach number $M_\infty$ and wall-to-recovery temperature ratio $T_w/T_r$.}
	\label{fig:cases}
\end{figure}

The DNS database plays the most pivotal role in the PIML approach. It is expected that the relevance of the training flows to prediction flow significantly affects the prediction performance. It is of interest to study the ML performances of different training flow databases and to assess the confidence of ML prediction \emph{a priori}. To investigate this issue, the Reynolds stresses for the higher Mach Case M8Tw053 are predicted using three different training databases of flows with lower Mach numbers, and corresponding performances are compared. The DNS data of the flow M8Tw053 is only used for comparing against. Such an arrangement is relevant to the scenario of real-world applications, since most existing data of either DNS or experiments are limited to low-Reynolds-number and low-Mach-number regimes. Moreover, to better examine the robustness of the PIML approach, the training is also conducted on the combined training datasets of all three lower-Mach-number flows. Specifically, the discrepancy model is trained by three training flows (Case M2p5Tw1, Case M6Tw025, and Case M6Tw076) separately for the first three tests and is trained on the combined flow datasets for the fourth test. These training flows are different in free-stream Mach numbers and wall-to-recovery temperature ratios. The scatter points of the four flows in the space of the two flow parameters are plotted in Fig.~\ref{fig:cases}.

To quantitatively evaluate the training and test performance, we use L2 norm of the difference between the data $z_d$ and ML-predicted value $z_p$ over the entire mesh as the error metric $\delta^{err}$, which is expressed as,
	\begin{equation}
	\label{eq:metric}
	\delta^{err} = \frac{1}{N}\sum_{i=1}^{N}(\frac{z_d^i - z_p^i}{\max\limits_{1 \leq i \leq N} |z_d^i|})^2,
	\end{equation}
where $N$ is the number of data points. The training performance of different training databases is evaluated by calculating the training errors and out-of-bag (OOB) errors, and all the details are reported in Table~\ref{tab:train} and Table~\ref{tab:oob} in Appendix~\ref{sec:appendx2}. Overall, the training errors of each discrepancy component are very small for all training datasets, indicating that the trained RF models can perfectly capture $\Delta R$ for every training flow. This also can be clearly seen by examining training errors on reconstructed turbulent stress components, most of which are less than $0.01\%$. The OOB error, which estimates the test error, is different for each discrepancy component and also varying in different training cases. In general, the estimated test errors of trained RF models by different databases are all satisfactory. The flow databases with higher Mach number have lower OOB errors and the estimated generalization performance of the anisotropy discrepancy is better than that of turbulent kinetic energy. Next, the prediction performance of each trained model is examined by comparing with the test flow datasets M8Tw053. 

\begin{figure}[htbp]
		\centering
		\includegraphics[width=0.5\textwidth]{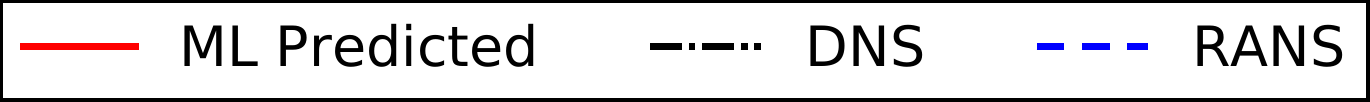}\\  
		\subfloat[$M = 6.0, T_w/T_r = 0.25$]{\includegraphics[width=0.25\textwidth]{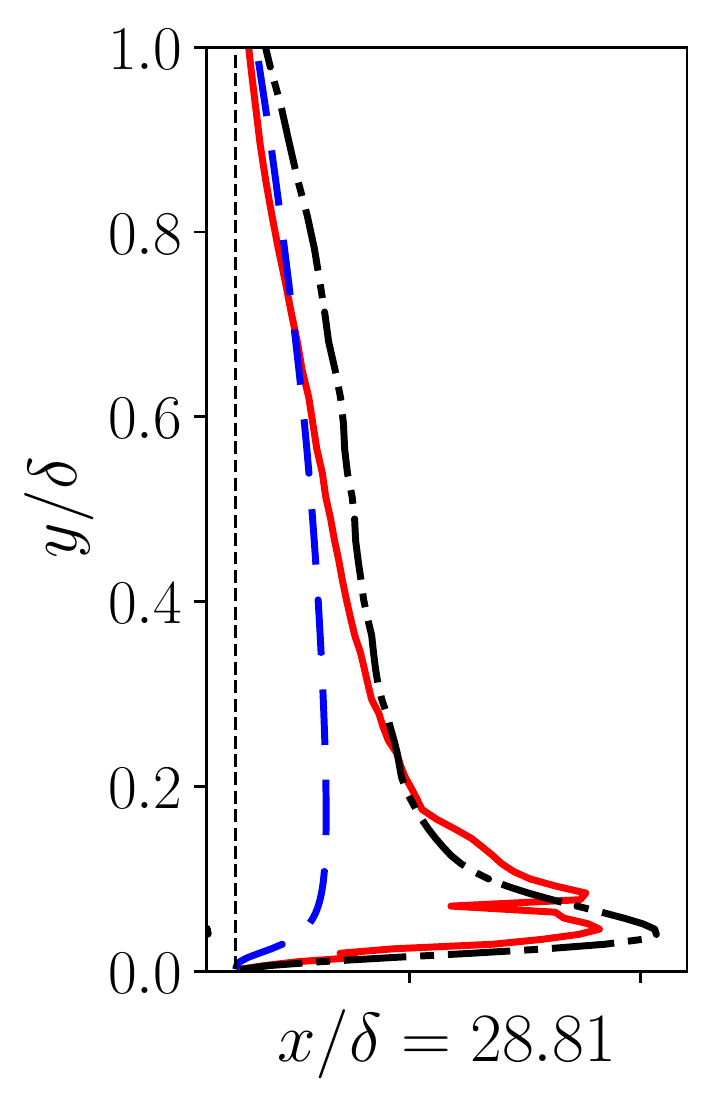}} 
		\subfloat[$M = 2.5, T_w/T_r = 1.0$]{\includegraphics[width=0.25\textwidth]{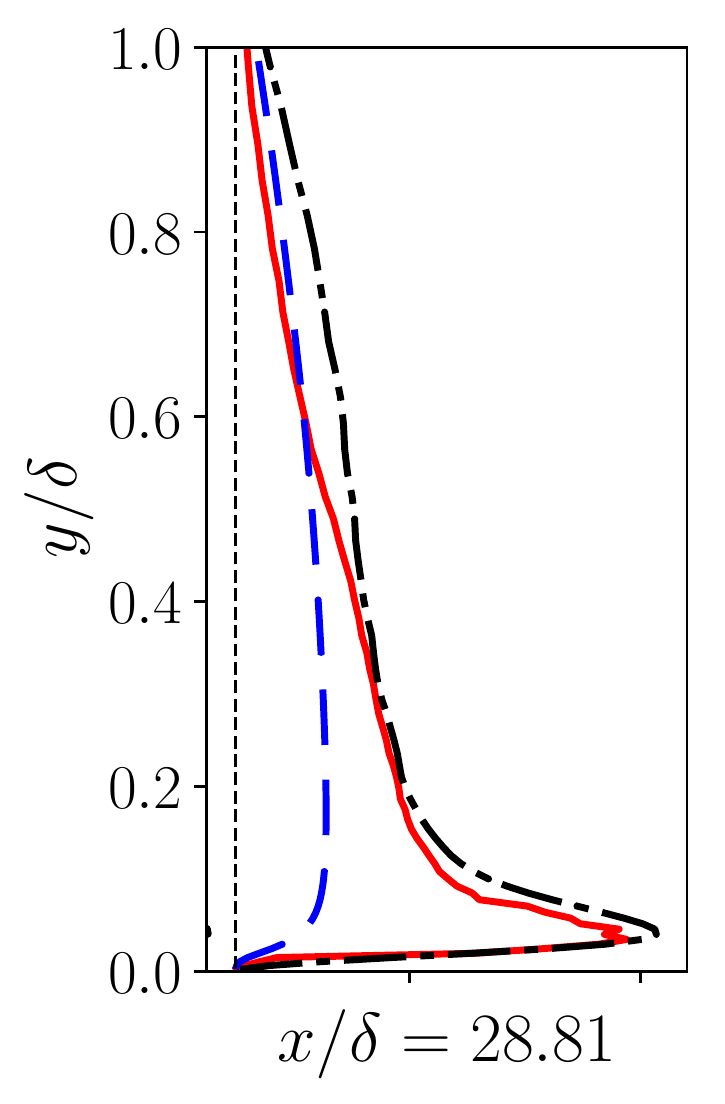}}
		\subfloat[$M = 6.0, T_w/T_r = 0.76$]{\includegraphics[width=0.25\textwidth]{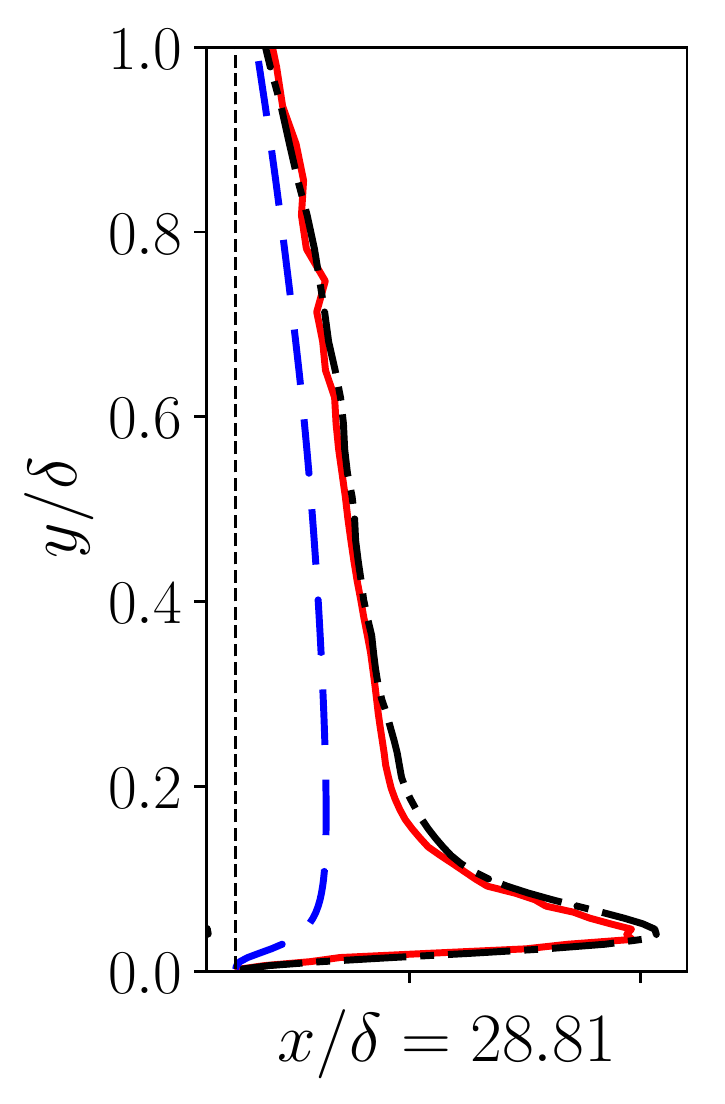}}
		\subfloat[Combined]{\includegraphics[width=0.25\textwidth]{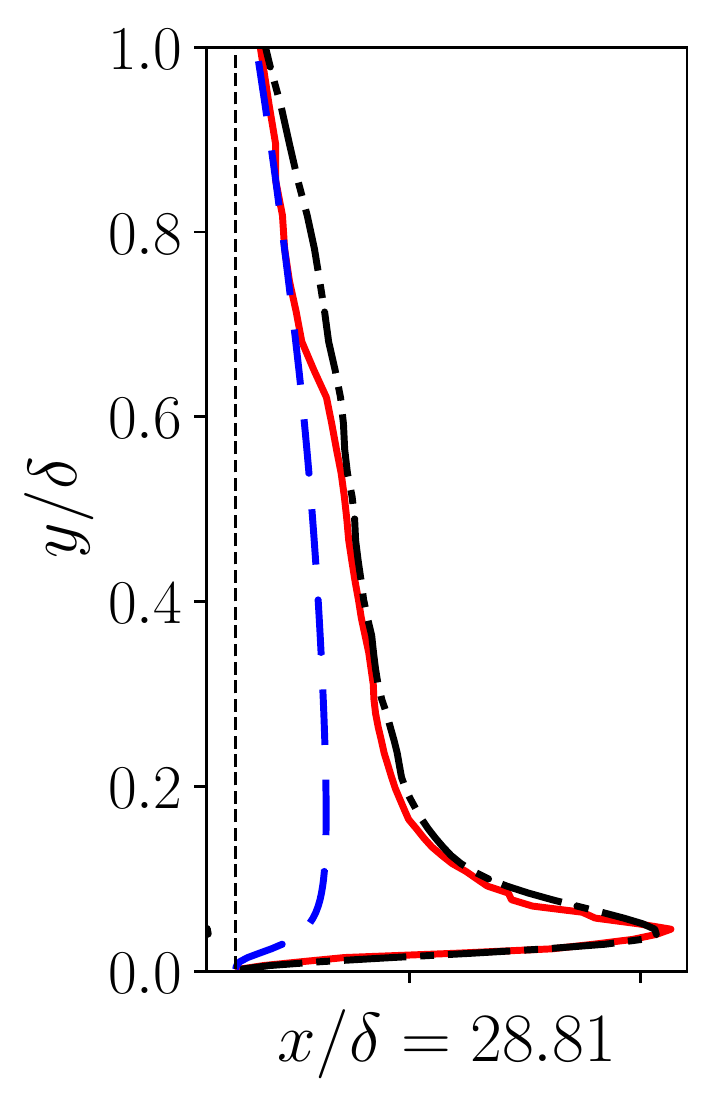}}		 		  
		\caption{Comparison of ML predicted turbulent normal stress $R_{xx} = \overline{\rho u' u'}$ learned from the training cases (a) M6Tw025, (b) M2p5Tw1, (c) M6Tw076, and (d) combined databases of M6Tw025, M2p5Tw1, and M6Tw076. Corresponding baseline RANS and DNS results are also plotted. Note that only the zoomed-in view of the profiles at $x/\delta = 28.81$ are shown.}
		\label{fig:compareRxx}
\end{figure}

Fig.~\ref{fig:compareRxx} compares the ML prediction performances on turbulent normal component $R_{xx}$ with the three aforementioned training flows and the combined datasets. The ML-predicted $R_{xx}$ profiles at $x/\delta = 28.81$ based on training flows M6Tw025, M2p5Tw1, M6Tw076, and combined flows are shown in Figs.~\ref{fig:compareRxx}a, \ref{fig:compareRxx}b, \ref{fig:compareRxx}c, and  \ref{fig:compareRxx}d, respectively. It can be seen that the ML-predicted $R_{xx}$ profiles in all training scenarios are corrected toward the DNS data. In the near-wall region, the peaks at $y/\delta = 0.1$, which are underestimated by the baseline RANS, are captured in all the four scenarios to different extents. Overall, the discrepancy (misfit) between RANS prediction and DNS data is reduced with ML corrections. However, the performance of the correction is notably different by using different training sets. When the flow M6Tw076 is used for training, the ML prediction is almost identical to the DNS data (Fig.~\ref{fig:compareRxx}c), since both the Mach number and the wall-to recovery temperature of the flow M6Tw076 are the closest to those of the prediction flow M8Tw053 in the three training flows. As the Mach number of the training flow is reduced to 2.5 (Fig.~\ref{fig:compareRxx}b), notable discrepancies can be found in the upper part of the boundary layer ($0.4< y/\delta < 1.0$). The wall temperature condition is also an important factor that affects closeness between training and prediction flows. Although the Mach number of flow M6Tw025 is close to the prediction flow, large bumps and non-smoothness are found in the near-wall region (Fig.~\ref{fig:compareRxx}a), since the training flow has a colder wall  ($T_w/T_r = 0.25$). This is different from the other training flows and the prediction flow, whose wall-to-recovery temperature ratios are all larger than 0.5. However, when training on the aggregated datasets of all three flows, the prediction performance of $R_{xx}$ significantly improved compared to the training cases M2p5Tw1 and M6Tw025 and is comparable to the case M6Tw076. It can be seen that the predicted $\Re_{xx}$ with combined training datasets agrees best with the DNS in the near-wall region. 

\begin{figure}[htbp]
	\centering
	\includegraphics[width=0.5\textwidth]{legend}\\ 
	\subfloat[$M = 6.0, T_w/T_r = 0.25$]{\includegraphics[width=0.25\textwidth]{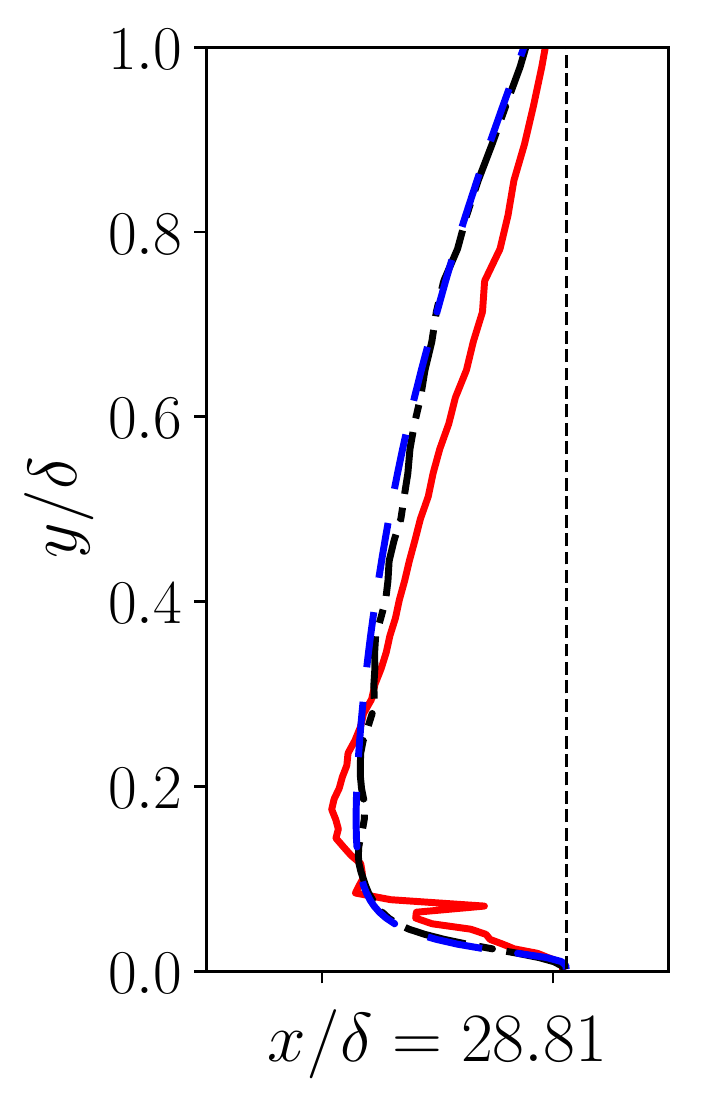}} 
	\subfloat[$M = 2.5, T_w/T_r = 1.0$]{\includegraphics[width=0.25\textwidth]{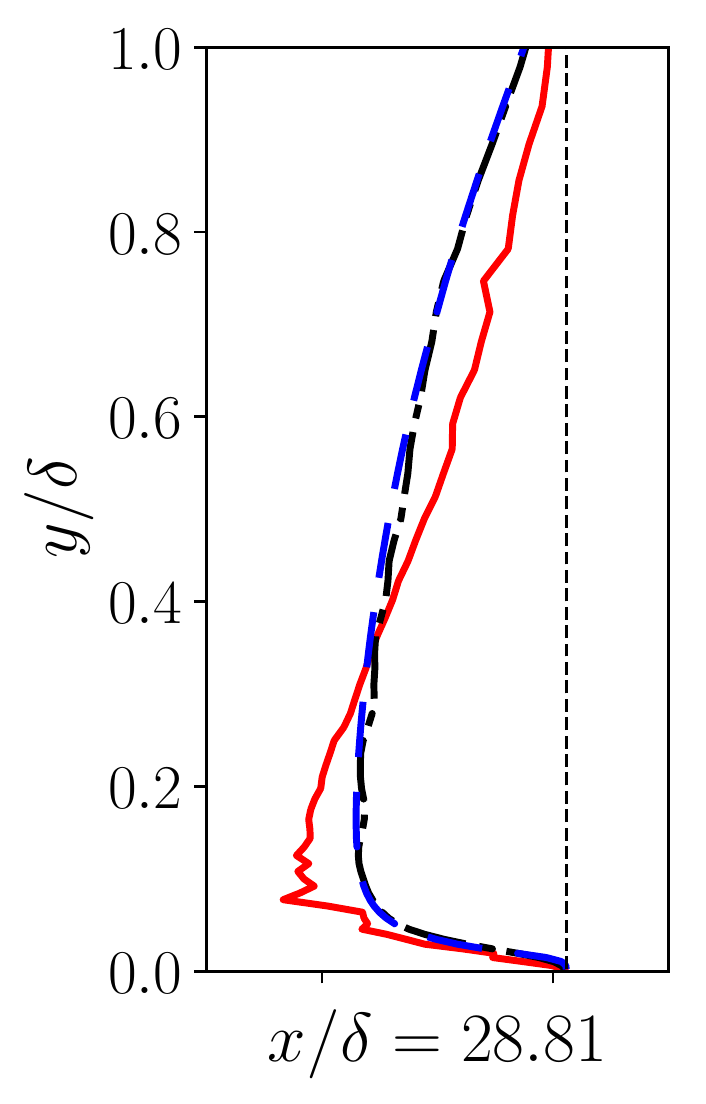}}
	\subfloat[$M = 6.0, T_w/T_r = 0.76$]{\includegraphics[width=0.25\textwidth]{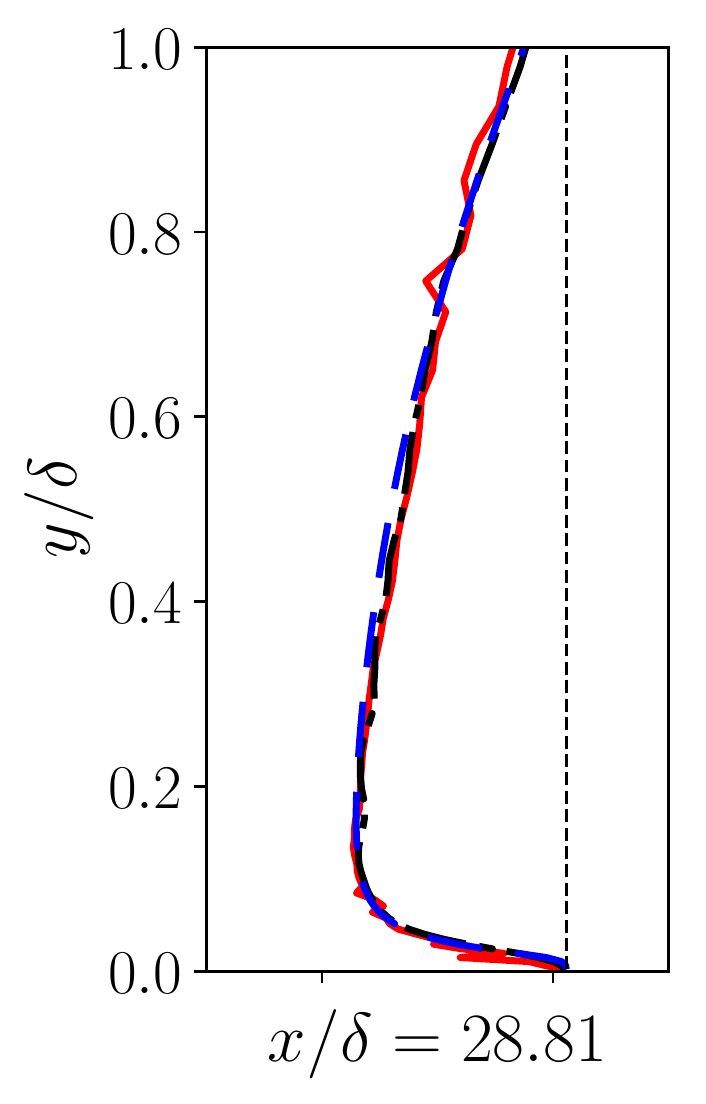}}
	\subfloat[Combined]{\includegraphics[width=0.25\textwidth]{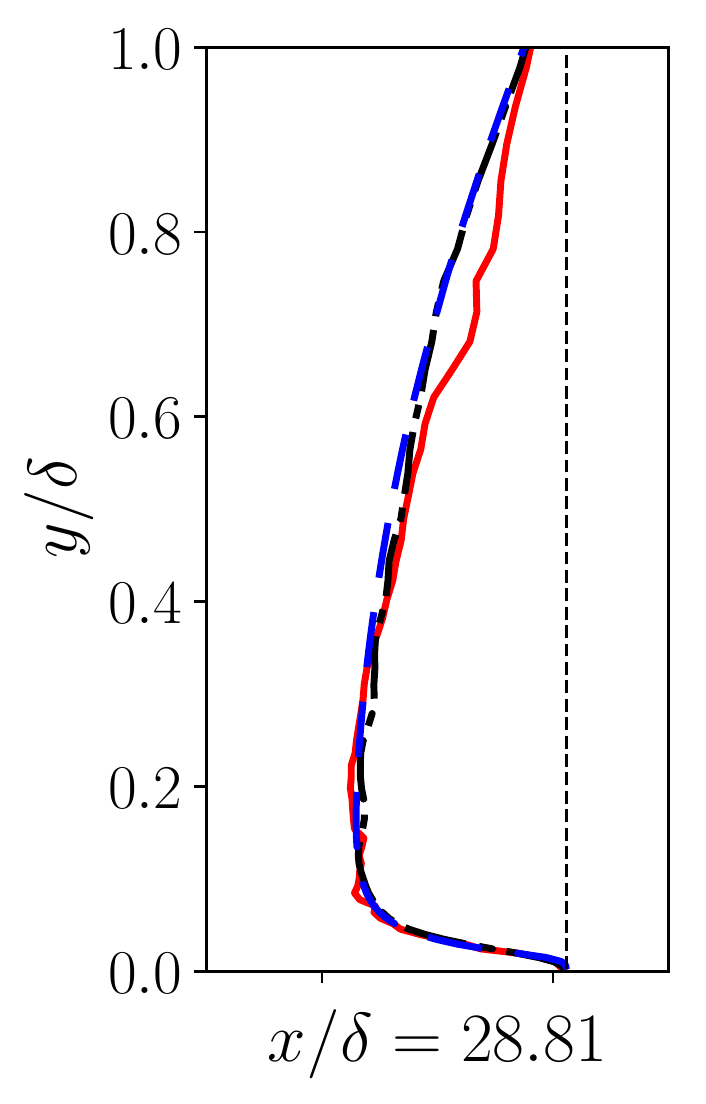}}			 		  
	\caption{Comparison of ML predicted turbulent shear stress $R_{xy} = \overline{\rho u' v'}$ learned from the training cases (a) M6Tw025, (b) M2p5Tw1, (c) M6Tw076, and (d) combined databases of M6Tw025, M2p5Tw1, and M6Tw076. Corresponding baseline RANS and DNS results are also plotted. Note that only the zoomed-in view of the profiles at $x/\delta = 28.81$ are shown.}
	\label{fig:compareRxy}
\end{figure}

The predictions of turbulent shear component $R_{xy}$ show the similar trend, and the comparisons are presented in Fig.~\ref{fig:compareRxy}. As mentioned in Sec.~\ref{sec:results}, the baseline RANS has been well-calibrated for turbulent shear stress, and thus different degrees of deterioration are observed in ML-predictions in all the three training scenarios. For the training cases with a low-Mach number (M2p5Tw1) or a cold-wall condition (M6Tw025), the ML-predicted $R_{xy}$ profiles deviate from the baseline and DNS results in the upper region of the boundary layer, and there are marked bumps near the wall. For the training case M6Tw076, the ML-prediction shows a better agreement with the DNS data. However, slight non-smoothness still can be observed, which is due to the pointwise estimation of the currently used machine learning algorithm as mentioned above. 
The non-smoothness is improved to a certain extent by using combined databases. In the near-wall-region, the ML-prediction is nearly identical to the DNS, showing more robustness.
In the PIML framework, the discrepancy of RANS-predicted Reynolds stress is learned and predicted by the ML model. When the discrepancy is already small (i.e., RANS prediction is accurate enough), errors associated with training data and ML learning process become more visible and thus the ML-corrected Reynolds stress may even deteriorate. It is wise to put less weight on the ML correction for regions where the RANS prediction performance is satisfactory. The RANS performance evaluation framework proposed by Ling et al.~\citep{ling2015evaluation} can be potentially employed to identify those regions and assign correction weights. This issue will be explored separately in the future work. 

To quantitatively evaluate the training-prediction performance of each case, we calculated the test errors for all predicted $\Delta R$ components and reconstructed Reynolds stress components based on the L2 error metric defined by Eq.~\ref{eq:metric}. The results are reported in Table.~\ref{tab:test}. 
\begin{table}[htbp]
	\centering
	\caption{Test errors of flow M8Tw053 using different training datasets}
	\label{tab:test} 
	\resizebox{\textwidth}{!}{\begin{tabular}{|l||*{5}{c|}}\hline
		\backslashbox[30mm]{Training database}{Test error (L2) in}
		&\makebox[3em]{$\Delta x_b$}&\makebox[3em]{$\Delta y_b$}&\makebox[3em]{$\Delta \mathrm{Log}(k)$}&\makebox[3em]{$\Delta \phi$}&\makebox[3em]{Total$^a$}\\\hline\hline
		M6Tw076 &7.40e-3&4.30e-3&7.94e-3&2.94e-2&1.58e-2\\\hline
		M6Tw025 &1.66e-2&5.41e-3&2.15e-1&4.41e-2&1.10e-1\\\hline
		M2p5Tw1 &3.18e-2&1.48e-2&2.34e-2&5.47e-2&3.38e-2\\\hline
		M6Tw076, M6Tw025, M2p5Tw1&4.10e-3&1.58e-3&1.39e-2&1.97e-2&1.23e-2\\\hline\hline
		\backslashbox[30mm]{Training database}{Test error (L2) in}
		&\makebox[3em]{$R_{xx}$}&\makebox[3em]{$R_{yy}$}&\makebox[3em]{$R_{zz}$}&\makebox[3em]{$R_{xy}$}&\makebox[3em]{Total$^a$}\\\hline\hline
		M6Tw076 &1.66e-3&3.74e-3&4.14e-3&3.97e-3&3.52e-3\\\hline
		M6Tw025 &1.11e-2&1.50e-2&1.90e-2&1.54e-2&1.54e-2\\\hline
		M2p5Tw1 &6.35e-3&2.66e-2&2.73e-2&2.45e-2&2.28e-2\\\hline
		M6Tw076, M6Tw025, M2p5Tw1&1.66e-3&2.44e-3&2.51e-3&2.79e-3&2.38e-3\\\hline\hline
		Baseline RANS error
		&\makebox[3em]{1.08e-1}&\makebox[3em]{2.89e-1}&\makebox[3em]{2.97e-1}
		&\makebox[3em]{1.09e-3}&\makebox[3em]{2.14e-1}\\\hline
	\end{tabular}}
	\flushleft
	\footnotesize{$^a$ Total error is defined as the mean square root of the errors of the first four columns.}
\end{table}
Overall, the test errors are significantly less than the baseline RANS error, especially for the turbulent normal stresses ($R_{xx}, R_{yy}, R_{zz}$). The accuracy of ML-corrected turbulent normal stresses has been improved by orders of magnitude. However, the error of the RANS-predicted turbulent shear stress ($R_{xy}$) is already very small, which is about two orders of magnitude smaller than those of the other normal stress components. As expected, the ML-corrected $R_{xy}$ deteriorates to a certain extent and the ML test errors are larger than that of the RANS-predicted shear stress component. Nonetheless, the accuracy of ML-corrected $R_{xy}$ by using either datasets M6Tw076 or combined datasets is comparable to that of the RANS prediction. By comparing the prediction performance of the three different training flows, we can see that the training with database M6Tw076 provides the most accurate prediction, the error of which is approximately two orders of magnitude smaller than that of the baseline RANS. It is worth to highlight that training on the combination of all three datasets produces the most accurate and robust prediction, where the test error is smaller than that of any cases with each individual training flow.

The results shown above demonstrate that the performance of the ML prediction largely depends on the data set used for training. Intuitively, the training flows M2p5Tw1 and M6Tw025 are less similar to the prediction flow M8Tw053 than the training flow M6Tw076 is, since they either have a much lower Mach number or have a colder wall. Therefore, their prediction performances deteriorate compared to that of the case M6Tw076, which is closer to the prediction flow in both Mach number and wall-temperature condition. A legitimate question is how to assess the prediction performance \emph{a priori} with the given database. This assessment is of great importance since it can provide guidelines to decide if the database is sufficient or not for training and which flow data should be added to improve prediction performance. Moreover, the assessment can identify the flow regions where the features are not well-supported by the training set, and so the ML correction can be masked in these areas. To quantitatively assess the ``closeness'' and ``difference'' between the training and predicted flow sets, Wu et al.~\citep{mfu9} proposed to use the RANS-simulated mean flow features. Specifically, RANS simulations can be performed for both the training and prediction flows to obtain the invariant mean flow features. The ``closeness'' of training and prediction flows can be assessed by visualizing the mean flow feature space of each flow and measuring the ``distance'' between training and prediction flows in the feature space.

\begin{figure}[htbp]
	\centering
	\includegraphics[width=0.7\linewidth]{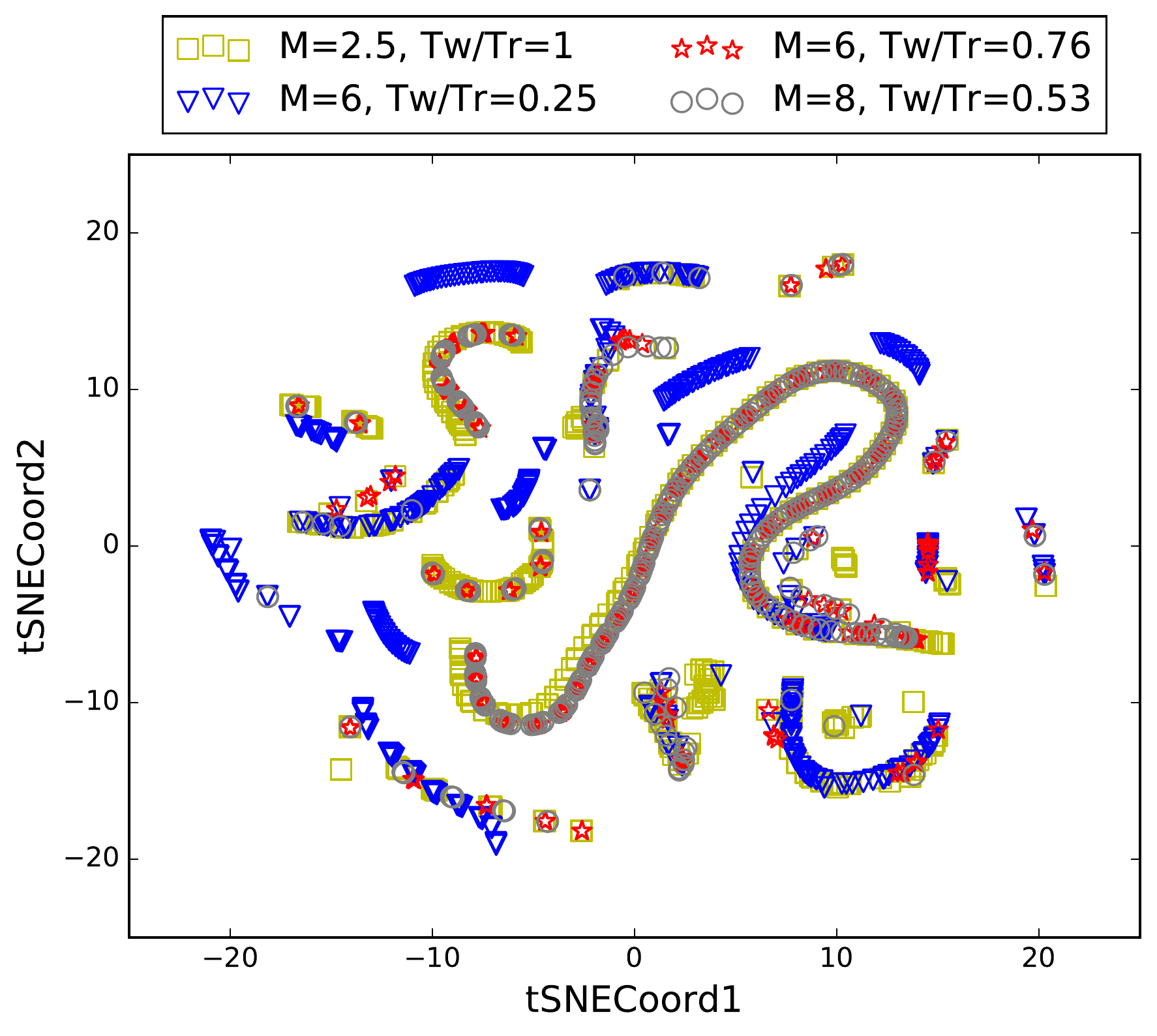}
	\caption{Two-dimensional t-SNE visualization of the local mean flow features of four flow cases M2p5Tw1, M6Tw025, M6Tw076, and M8Tw053. The two t-SNE coordinates are denoted by tSNECood1 and tSNECood2, respectively.}
	\label{fig:tSNE}
\end{figure}

First, it is of interest to visualize training and prediction flows in mean feature space. However, this is not trivial since each point in the feature space has 47 dimensions. Dimension reduction techniques are needed to interpret the high-dimensional feature points visually. The t-Distributed Stochastic Neighbor Embedding (t-SNE) technique~\citep{maaten2008visualizing} is specifically designed to enable the visualization of high-dimensional data sets by mapping the high-dimensional data into a two- or three-dimensional space. To identify a low-dimensional manifold with the pairwise similarity of the high-dimensional space, the conditional probabilities converted from Euclidean distances between data points in both high- and low-dimensional spaces are calculated. The cost function based on Kullback-Leibler divergence of the two probabilities is minimized to find the best representation of high-dimensional data in a low-dimensional manifold. 
Refer to Maaten et al.~\citep{maaten2008visualizing} for more details of t-SNE. Wu et al.~\citep{wu2017visualization} proposed using t-SNE to visualize high dimensional turbulence simulation data. Here, we employ t-SNE dimension reduction technique to visually analyze the closeness of different training flows to the prediction flow. Fig.~\ref{fig:tSNE} shows the mean flow feature points of the cases, M2p5Tw1, M6Tw025, M6Tw076, and M8Tw053, in two-dimensional t-SNE coordinates. It can be clearly seen that the points of flow M6Tw076 (stars) align well with those of the prediction flow M8Tw053 (circles). The points of both flows M2p5Tw1 (squares) and M6Tw025 (triangles) deviate from those of the prediction flow. Particularly for the cold-wall case (M6Tw025), most t-SNE points are located far away from those of the prediction flow, indicating that the prediction case is not supported by the cold-wall case in most regions. After mapping back the t-SNE points that are not supported by the training case of M6Tw025 to the physical coordinates, we found that these unsupported points are located near the wall ($y/\delta < 0.15$) in the physical domain. This region is the one where large unphysical bumps exist in the ML-predictions with training flow M6Tw025 (see Figs.~\ref{fig:compareRxx}a and~\ref{fig:compareRxy}a). The t-SNE visualization for the training and prediction flows is consistent with the prediction performance of each training flow.

In additional to tSNE that provides a qualitative visualization of the regions unsupported by training flows, the Mahalanobis distance (M-distance) $D_m$ is employed to quantitatively measure the closeness between training and prediction flows and explain the different prediction performances shown above according to Wu et al.~\citep{mfu9}. The M-distance is defined as the distance between a feature point $\mathbf{q}$ of the prediction flow and the mean value $\bs{\mu}$ of the feature points of the training flow, scaled by the covariance matrix $\Sigma$ of all the training points, 
\begin{equation}
D_m = \sqrt{(\mathbf{q} - \bs{\mu})^T \Sigma^{-1}(\mathbf{q} - \bs{\mu})}.
\end{equation}
The M-distance $D_m$ is normalized to the range between 0 to 1 based on percentiles from the training set. Larger M-distance means higher degree of extrapolation from training to prediction flows. 

\begin{figure}[htbp]
	\centering
	\includegraphics[width=0.7\linewidth]{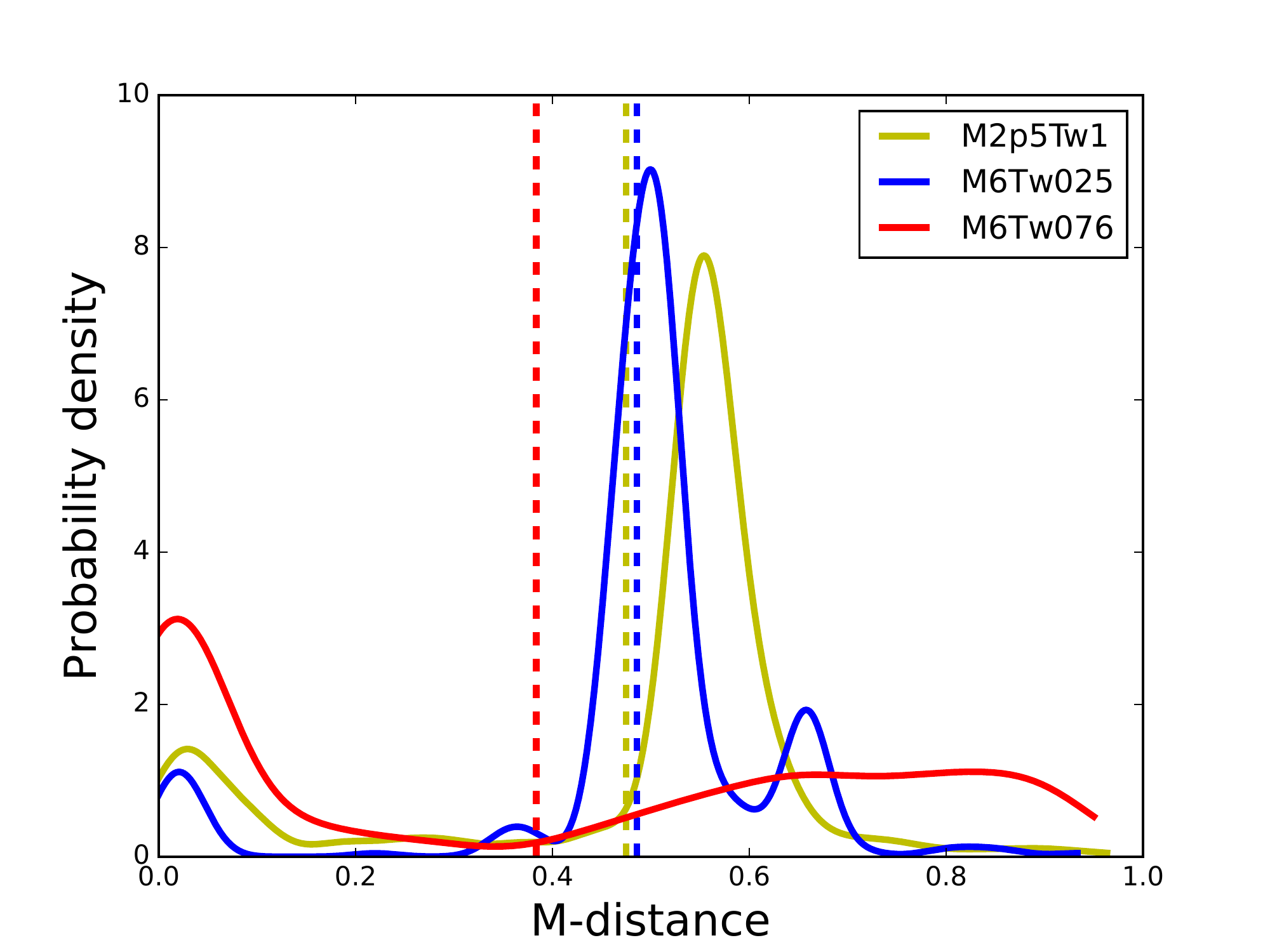}
	\caption{Probability density function of Mahalanobis distance based on different training sets. All the distances have been normalized into the range from zero to one. }
	\label{fig:Mdist}
\end{figure}

Fig.~\ref{fig:Mdist} shows distributions of M-distances from the prediction flow to the three training sets, where the mean value of the M-distance of each training case is also plotted. As expected, most points of prediction flow have small M-distances to the training flow M6Tw076, and the mean M-distance is 0.38. This is consistent with the intuition since the Mach number and wall temperature ratio are close to the prediction flow in this training scenario. The M-distances to training flows M6Tw025 and M2p5Tw1 become larger, and the distances are greater than 0.45 for most points. The mean M-distances for M6Tw025 and M2p5Tw1 are significantly larger than that of case M6Tw076. It can be seen that the averaged M-distances of the cold-wall case is slightly larger than that of low-Mach number case. The observations in Fig.~\ref{fig:Mdist} can be well correlated to the prediction performances shown in Figs.~\ref{fig:R_xx_M25} and~\ref{fig:R_xy_M25}. Moreover, the M-distance of each point in the prediction flow can be used to evaluate the prediction confidence and to mask the prediction quantitatively.

\section{Summary}
\label{sec:con}
In this paper, a physics-informed machine learning (PIML) approach is applied to improve the accuracy of the RANS-modeled Reynolds stresses for high-speed flat-plate turbulent boundary layers using an existing DNS database. The effectiveness of the PIML technique for improving Reynolds stresses is demonstrated in several training scenarios where the training and prediction datasets differ both in the freestream Mach number and in the wall-to-recovery temperature ratio. The study shows that the RANS-modeled turbulent normal stresses, the turbulent kinetic energy, and the Reynolds-stress anisotropy can be significantly improved using the PIML technique. Moreover, the learning-prediction performances of different training scenarios are also evaluated qualitatively by using a high-dimensional visualization technique t-SNE and quantitatively by using a distance metric, M-distance. Both techniques provide a priori assessment of the prediction confidence with a given training data set. The PIML methodology consists of an economical technique for improving the accuracy of RANS-modeled Reynolds stresses for high-speed turbulent flows when there is a lack of experiments or high-fidelity simulations.

Although the main aim in this work is to reduce the model-form error of the RANS-predicted Reynolds stress, ultimate goal of the PIML framework is to improve the RANS-predicted mean flow field by solving the PDE with the ML-corrected Reynolds stress. Success has been achieved in a number of incompressible flows~\citep{wang2016physics,mfu14,Wu2016}. but the significant challenge still remains. The primary limitation is due to the non-smoothness observed in certain ML predictions, because it is the divergence of the Reynolds stress appears in the RANS equations. Namely, despite the fact that the ML predictions are closer to the truth, the non-smooth Reynolds stress field (i.e., slight wiggling around the truth) might pollute the propagated velocity field. There are two main reasons for this issue. First, the machine learning is conducted pointwisely without consideration of spatial correlation information. Therefore, smooth predictions and improved spatial derivatives of Reynolds stresses may not necessarily be guaranteed. Possible solutions could include imposing the spatial correlation constraint on the learning process or developing other learning-prediction strategies which do not depend on pointwise training. Second, it is known that the prediction performance is highly related to the quality and quantity of the training data. For the cases where features of the test flow cannot be well supported by those in the training flows (e.g., cold-wall training set M6Tw025), non-smoothness is more likely to happen. The prediction performance assessment methods presented in this work can potentially be utilized to identify the flow regions where the test errors might be large \emph{a priori}, and the ML-corrections in such regions can be less weighted accordingly. On the other hand, including more relevant datasets for training will also lend robustness and obtain a more smooth and accurate prediction.

In this work, a basis of 47 input invariants is constructed in the current PIML framework, but a considerably smaller number of input invariants might be sufficient to model high-speed flat-plate turbulent boundary layers, given that the direct compressibility effects remain small even for Mach numbers up to 12~\citep{duan2011direct} and, based on the Morkovin's hypothesis, there are simple physics based incompressible-compressible transformations such as those derived by Huang et al.~\citep{huang1995compressible} and Trettel and Larsson~\citep{trettel2016mean}. Therefore, the current study should be seen as a first step to extend the PIML framework by Wang et al.~\citep{Wang2016}. to high-speed compressible flows. Further study is required for deriving machine learning approaches that are better informed by physics in the high-speed regime and for reducing the dimensional space of the machine-learning algorithm. The modeling and scaling of turbulent statistics as given in Huang et al.~\citep{huang1995compressible} and Trettel and Larsson~\citep{trettel2016mean} should provide a good starting point for deriving optimized ``physics informed'' approaches. Moreover, the dimension of the feature space can be further reduced by analyzing the importance of each feature \textit{a posteriori}. For example, the feature importance measure in the Random Forest can be studied to identify the critical features most relevant to construct the Reynolds stress discrepancy~\citep{Wang2016}.

\begin{acknowledgements}
The DNS database was produced based upon the work supported by AFOSR under Grant FA9550-14-1-0170 (Program Manager I. Leyva) and NASA Langley Research Center under Grant NNL09AA00A (through the National Institute of Aerospace). Computational resources for the DNS were provided by the NASA Advanced Supercomputing Division, the DoD High Performance Computing Modernization Program, and the NSF's Petascale Computing Resource Allocations Program (NSF ACI-1640865). Any opinions, findings, and conclusions or recommendations expressed in this material are those of the authors and do not necessarily reflect the views of the United States Air Force. We also thank the anonymous reviewers for their comments, which helped improving the quality and clarity of the manuscript.
\end{acknowledgements}


\section*{Appendix 1: Integrity Basis of Mean Flow Features of High-Speed Flows}
\label{sec:appendx1}
The minimal integrity invariance bases for mean-flow vectors and tensors of high-speed flows are given in Table~\ref{tab:basis}. Note that the vectors $\nabla T$ and $\nabla k$ should be first mapped to antisymmetric tensors
	as follows,
	\begin{subequations}
		\label{eq:vector2anti}
		\begin{align}
		\widehat{\mathbf{A}}_T & =  -\mathbf{I} \times \widehat{\nabla T}\\
		\widehat{\mathbf{A}}_k & =  -\mathbf{I} \times \widehat{\nabla k}
		\end{align}  	 
	\end{subequations}
where $\mathbf{I}$ is the second order identity tensor, and $\times$ denotes tensor cross product. The asterisk ($*$) on a term means to include all terms formed by cyclic permutation of anti-symmetric tensor labels (e.g., $\widehat{\bs{\Omega}}^2 \widehat{\mathbf{A}}_{T} \widehat{\mathbf{S}}$* is short for $\widehat{\bs{\Omega}}^2 \widehat{\mathbf{A}}_{T} \widehat{\mathbf{S}}$ and $ \widehat{\mathbf{A}}_{T}^2 \widehat{\bs{\Omega}} \widehat{\mathbf{S}}$). 
\begin{table}[htbp]  
	\centering
	\caption{Minimal integrity bases for symmetric tensor $\widehat{\mathbf{S}}$ and antisymmetric tensors $\widehat{\bs{\Omega}}$, $\widehat{\mathbf{A}}_{T}$, and $\widehat{\mathbf{A}}_{k}$. In the implementation, $\widehat{\mathbf{S}}$ is the rate of strain tensor, $\widehat{\bs{\Omega}}$ is the rate of rotation tensor; $\widehat{\mathbf{A}}_{T}$ and $\widehat{\mathbf{A}}_{k}$ are the antisymmetric tensors associated with temperature gradient $\widehat{\nabla T}$ and the gradient of turbulent kinetic energy $\widehat{\nabla k}$. $n_S$ and $n_A$ denote the numbers of symmetric and antisymmetric raw tensors for the bases; an asterisk ($*$) on a term means to include all terms formed by cyclic permutation of labels of anti-symmetric tensors. Note the invariant bases are traces of the tensors in the third column. }
	\label{tab:basis}
	\resizebox{\textwidth}{!}{\begin{tabular}{c|C{2.5cm}|C{9.5cm}}	
		\hline
		$(n_S, n_A)$ &  feature index &  invariant bases$^{(\mathrm{a})}$\\
		\hline
		(1, 0) & 1--2 & $\widehat{\mathbf{S}}^2$, $\widehat{\mathbf{S}}^3$ \\
		\hline
		(0, 1)& 3--5 & $\widehat{\bs{\Omega}}^2$,  $\widehat{\mathbf{A}}_{T}^2$,  $\widehat{\mathbf{A}}_{k}^2$ \\
		\hline
		\multirow{3}{*}{(1, 1)} & \multirow{3}{*}{6--14} 
		& $\widehat{\bs{\Omega}}^2 \widehat{\mathbf{S}}$, $\widehat{\bs{\Omega}}^2 \widehat{\mathbf{S}}^2$, $\widehat{\bs{\Omega}}^2 \widehat{\mathbf{S}} \widehat{\bs{\Omega}} \widehat{\mathbf{S}}^2$;\\
		&& $\widehat{\mathbf{A}}_{T}^2 \widehat{\mathbf{S}}$, $\widehat{\mathbf{A}}_{T}^2 \widehat{\mathbf{S}}^2$, $\widehat{\mathbf{A}}_{T}^2 \widehat{\mathbf{S}} \widehat{\mathbf{A}}_{T} \widehat{\mathbf{S}}^2$;\\
		&& $\widehat{\mathbf{A}}_{k}^2 \widehat{\mathbf{S}}$, $\widehat{\mathbf{A}}_{k}^2 \widehat{\mathbf{S}}^2$, $\widehat{\mathbf{A}}_{k}^2 \widehat{\mathbf{S}} \widehat{\mathbf{A}}_{k} \widehat{\mathbf{S}}^2$; \\
		\hline
		(0, 2)& 15--17 & $\widehat{\bs{\Omega}} \widehat{\mathbf{A}}_{T}$, $\widehat{\mathbf{A}}_{T} \widehat{\mathbf{A}}_{k}$, $\widehat{\bs{\Omega}} \widehat{\mathbf{A}}_{k}$ \\
		\hline
		\multirow{3}{*}{(1, 2)} & \multirow{3}{*}{18--41} & 
		$\widehat{\bs{\Omega}} \widehat{\mathbf{A}}_{T} \widehat{\mathbf{S}}$, $\widehat{\bs{\Omega}} \widehat{\mathbf{A}}_{T} \widehat{\mathbf{S}}^2$, $\widehat{\bs{\Omega}}^2 \widehat{\mathbf{A}}_{T} \widehat{\mathbf{S}}$*, $\widehat{\bs{\Omega}}^2 \widehat{\mathbf{A}}_{T} \widehat{\mathbf{S}}^2$*, $\widehat{\bs{\Omega}}^2 \widehat{\mathbf{S}} \widehat{\mathbf{A}}_{T} \widehat{\mathbf{S}}^2$*;\\
		
		&&$\widehat{\bs{\Omega}} \widehat{\mathbf{A}}_{k} \widehat{\mathbf{S}}$,  $\widehat{\bs{\Omega}} \widehat{\mathbf{A}}_{k} \widehat{\mathbf{S}}^2$,
		$\widehat{\bs{\Omega}}^2 \widehat{\mathbf{A}}_{k} \widehat{\mathbf{S}}$*,  $\widehat{\bs{\Omega}}^2 \widehat{\mathbf{A}}_{k} \widehat{\mathbf{S}}^2$*, $\widehat{\bs{\Omega}}^2 \widehat{\mathbf{S}} \widehat{\mathbf{A}}_{k} \widehat{\mathbf{S}}^2$*;\\
		
		&&$\widehat{\mathbf{A}}_{T} \widehat{\mathbf{A}}_{k} \widehat{\mathbf{S}}$, $\widehat{\mathbf{A}}_{T} \widehat{\mathbf{A}}_{k} \widehat{\mathbf{S}}^2$,
		$\widehat{\mathbf{A}}_{T}^2 \widehat{\mathbf{A}}_{k} \widehat{\mathbf{S}}$*, $\widehat{\mathbf{A}}_{T}^2 \widehat{\mathbf{A}}_{k} \widehat{\mathbf{S}}^2$*, $\widehat{\mathbf{A}}_{T}^2 \widehat{\mathbf{S}} \widehat{\mathbf{A}}_{k} \widehat{\mathbf{S}}^2$*;\\
		\hline
		(0, 3) & 42 & $\widehat{\bs{\Omega}} \widehat{\mathbf{A}}_{T} \widehat{\mathbf{A}}_{k}$ \\
		\hline
		(1, 3) & 43--47 &   $\widehat{\bs{\Omega}} \widehat{\mathbf{A}}_{T} \widehat{\mathbf{A}}_{k} \widehat{\mathbf{S}}$,  $\widehat{\bs{\Omega}} \widehat{\mathbf{A}}_{k} \widehat{\mathbf{A}}_{T} \widehat{\mathbf{S}}$,  $\widehat{\bs{\Omega}} \widehat{\mathbf{A}}_{T} \widehat{\mathbf{A}}_{k} \widehat{\mathbf{S}}^2$,
		$\widehat{\bs{\Omega}} \widehat{\mathbf{A}}_{k} \widehat{\mathbf{A}}_{T} \widehat{\mathbf{S}}^2$,  $\widehat{\bs{\Omega}} \widehat{\mathbf{A}}_{T} \widehat{\mathbf{S}} A_3 \mathbf{S}^2$ \\
		\hline						 								
	\end{tabular}}
	\flushleft
	{\small
		Note: (a) The invariance basis is the trace of each tensor listed below.}	
\end{table}

\section*{Appendix 2: Training Performance of Different Training Databases}
\label{sec:appendx2}

The training errors and out-of-bag (OOB) errors are clcalculated on both Reynolds stress discrepancy components (i.e., $\Delta R = [\Delta x_b, \Delta y_b, \Delta\log(k), \Delta\phi]$) and reconstructed turbulent stress components (i.e., $R_{xx}, R_{yy}, R_{zz}, R_{xy}$). Note that turbulent shear stresses $R_{xz}$ and $R_{yz}$ are negligible in boundary layer flows. The calculated training errors and OOB are listed by Table.~\ref{tab:train} and Table.~\ref{tab:oob}, respectively.

\begin{table}[htbp]
	\centering
	\caption{Training errors of different training datasets
	}
	\label{tab:train}
	\resizebox{\textwidth}{!}{\begin{tabular}{|l||*{5}{c|}}\hline
		\backslashbox[10mm]{Training database}{Training error (L2) in}
		&\makebox[3em]{$\Delta x_b$}&\makebox[3em]{$\Delta y_b$}&\makebox[3em]{$\Delta \mathrm{Log}(k)$}&\makebox[3em]{$\Delta \phi$}&\makebox[3em]{Total$^a$}\\\hline\hline
		M8Tw053 &6.76e-5&4.31e-5&1.07e-5&1.44e-4&8.26e-5\\\hline
		M6Tw076 &9.39e-5&8.32e-5&1.46e-5&1.73e-4&1.07e-4\\\hline
		M6Tw025 &9.29e-5&5.35e-5&1.03e-3&5.07e-4&5.77e-4\\\hline
		M2p5Tw1 &7.83e-5&3.72e-5&6.54e-5&1.82e-3&9.12e-4\\\hline\hline
		\backslashbox[10mm]{Training database}{Training error (L2) in}
		&\makebox[3em]{$R_{xx}$}&\makebox[3em]{$R_{yy}$}&\makebox[3em]{$R_{zz}$}&\makebox[3em]{$R_{xy}$}&\makebox[3em]{Total$^a$}\\\hline\hline
		M8Tw053 &1.05e-5&2.37e-5&2.20e-5&8.64e-5&4.64e-5\\\hline
		M6Tw076 &4.48e-6&1.97e-5&2.16e-5&6.22e-5&3.44e-5\\\hline
		M6Tw025 &8.37e-4&7.04e-4&6.42e-4&1.86e-3&1.12e-3\\\hline
		M2p5Tw1 &7.95e-6&2.24e-5&2.03e-5&4.09e-5&2.57e-5\\\hline
	\end{tabular}}
	\flushleft
	\footnotesize{$^a$ Total error is defined as the mean square root of the errors in the first four columns.}
\end{table}

\begin{table}[htbp]
	\centering
	\caption{Out-of-bag (OOB) errors of different training datasets}
	\label{tab:oob}
	\resizebox{\textwidth}{!}{\begin{tabular}{|l||*{5}{c|}}\hline
		\backslashbox[10mm]{Training database}{OOB errors in}
		&\makebox[3em]{$\Delta x_b$}&\makebox[3em]{$\Delta y_b$}&\makebox[3em]{$\Delta \mathrm{Log}(k)$}&\makebox[3em]{$\Delta \phi$}&\makebox[4em]{Averaged$^a$}\\\hline\hline
		M8Tw053 &3.05e-3&4.46e-3&8.74e-3&8.93e-2&2.64e-2\\\hline
		M6Tw076 &4.79e-3&1.03e-2&1.46e-3&1.41e-2&7.66e-3\\\hline
		M6Tw025 &7.60e-2&4.82e-2&6.51e-2&2.34e-2&5.32e-2\\\hline
		M2p5Tw1 &4.66e-2&9.78e-2&2.74e-2&2.57e-2&4.94e-2\\\hline
	\end{tabular}}
	\flushleft
	\footnotesize{$^a$ Averaged OOB errors of the first four columns.}
\end{table}

\end{document}